\documentclass[twocolumn,aps,pra,superscriptaddress]{revtex4-2}
\usepackage[T1]{fontenc}
\usepackage[utf8]{inputenc}
\usepackage{graphicx}
\usepackage{subfigure}
\usepackage{epsfig}
\usepackage{adjustbox}
\usepackage{amssymb,amsmath,amsopn,amsfonts}
\usepackage{colordvi}
\usepackage{color}
\usepackage{subeqnarray}
\usepackage{booktabs,tabularx}
\usepackage{siunitx}
\usepackage{bbold}
\usepackage[table]{xcolor}
\usepackage{siunitx}
\usepackage{braket}
\usepackage{bbold}
\usepackage[hidelinks]{hyperref}

\makeatletter
\def\p@subsection{}
\makeatother

\newcommand{\sech}{\mathrm{sech}}

\begin{document}

\title{Heralded generation of entanglement with photons}

\author{Imogen Forbes}
\affiliation{Quantum Engineering and Technology Laboratories, School of Physics and Department of Electrical and Electronic Engineering, University of Bristol, Bristol, BS8 1TL, United Kingdom}

\author{Farzad Ghafari}
\email{f.ghafari@griffith.edu.au}
\affiliation{Centre for Quantum Dynamics, Griffith University, Yuggera Country, Brisbane, QLD, 4111, Australia}

\author{Edward C. R. Deacon}
\affiliation{Quantum Engineering and Technology Laboratories, School of Physics and Department of Electrical and Electronic Engineering, University of Bristol, Bristol, BS8 1TL, United Kingdom}

\author{Sukhjit P. Singh}
\affiliation{Centre for Quantum Dynamics, Griffith University, Yuggera Country, Brisbane, QLD, 4111, Australia}

\author{Emilien Lavie}
\affiliation{Quantum Engineering and Technology Laboratories, School of Physics and Department of Electrical and Electronic Engineering, University of Bristol, Bristol, BS8 1TL, United Kingdom}

\author{Patrick Yard}
\affiliation{Quantum Engineering and Technology Laboratories, School of Physics and Department of Electrical and Electronic Engineering, University of Bristol, Bristol, BS8 1TL, United Kingdom}

\author{Reece D. Shaw}
\affiliation{Quantum Engineering and Technology Laboratories, School of Physics and Department of Electrical and Electronic Engineering, University of Bristol, Bristol, BS8 1TL, United Kingdom}

\author{Anthony Laing}
\affiliation{Quantum Engineering and Technology Laboratories, School of Physics and Department of Electrical and Electronic Engineering, University of Bristol, Bristol, BS8 1TL, United Kingdom}

\author{Nora Tischler}
\email{n.tischler@griffith.edu.au}
\affiliation{Centre for Quantum Dynamics, Griffith University, Yuggera Country, Brisbane, QLD, 4111, Australia}

\date{\today}
\begin{abstract}
Entangled states of photons form the backbone of many quantum technologies. 
Due to the lack of effective photon-photon interactions, the generation of these states is typically probabilistic.  
In the prevailing but fundamentally limited generation technique, known as postselection, the target photons are measured destructively in the generation process. 
By contrast, in the alternative approach---heralded state generation---the successful creation of a desired state is verified by the detection of ancillary photons.
Heralded state generation is superior to postselection in several critical ways: It enables free usage of the prepared states, allows for the success probability to be arbitrarily increased via multiplexing, and provides a scalable route to quantum information processing using photons.
Here, we review theoretical proposals and experimental realizations of heralded entangled photonic state generation, as well as the impact of realistic experimental errors.
We then discuss the wide-ranging applications of these states for quantum technologies, including resource states in linear optical quantum computing, entanglement swapping for repeater networks, fundamental physics, and quantum metrology.
\end{abstract}

\maketitle

\section{Introduction}

\subsection{Probabilistic generation of entanglement with photons}
Quantum entanglement between remote systems represents one of the sharpest divides between quantum and classical descriptions, underpinning many proposed applications of quantum mechanics. 
Entanglement is known to act as a resource in quantum computing~\cite{vidal2003entanglement}, device-independent quantum key distribution~\cite{briegel1998,liao2018,zukowski1993,wehner2018}, super dense coding \cite{hu2018beating}, and in quantum metrology~\cite{giovannetti2011advances,huang2016usefulness}.
From a hardware perspective, photons have emerged as a leading quantum information carrier in many technologies.
They suffer very little decoherence, allowing them to encode quantum information even at room temperature, and can be transmitted through low-loss optical fibres over large distances.
A prominent method for encoding qubits into photonic states, known as \textit{dual-rail encoding}, defines the logical qubit states by single excitations in one of two orthogonal modes as

\begin{equation}
    \ket{0} = \ket{10}_{F} = \hat{a}^\dagger_0\ket{00}_{F} \qquad  \ket{1} = \ket{01}_{F} = \hat{a}^\dagger_1\ket{00}_{F},
    \label{eq:def_dual_rail}
\end{equation}
where $\hat{a}^\dagger_i$ creates a photon in the $i^\mathrm{th}$ mode, and the subscript distinguishes Fock states from logical qubit states. 
The orthogonal modes can be realised in any of the photon's degrees of freedom such as polarisation, arrival time, orbital angular momentum, spatial mode, or frequency \cite{Crespi2011, Marcikic_2004, Giovannini_2011, OBrien2003, Clementi2023}.
In free space and fibre optics experiments, polarisation is often favoured due to the ease of its manipulation, with spatial mode degeneracy ameliorating alignment requirements. 
In the integrated photonics platform, waveguides that can be easily etched into a material define spatial modes. 
Dual-rail qubits have two distinct advantages, both stemming from the fact that the qubit states have a fixed total photon number. 
Firstly, single-qubit gates can be implemented deterministically using passive linear optics, and secondly, any photon loss is now flagged. If less photons than the number of qubits are detected, the specific location of the loss event is known.
This information can then be used to design loss-tolerant protocols. 

A significant drawback of dual-rail encoding stems from the lack of strong photon-photon interactions, which renders the generation of entanglement between qubits probabilistic. 
In linear optics, approaches for implementing arbitrary linear optical unitaries exist \cite{clements_optimal_2016,reck_experimental_1994}. 
However, using linear optical unitaries to implement arbitrary two-qubit operations, required for entanglement generation, results in occupation outside the computational basis states defined in Eq.~(\ref{eq:def_dual_rail}).

Two avenues exist for overcoming this limitation, with the first aiming to use the interaction of light and matter to mediate entanglement between photons. 
This can be in the form of Kerr interactions in nonlinear materials, using two-level systems such as NV centres, molecules, or quantum dots either to interact two incoming photons or to generate entangled photons directly through correlated emission. 
The second approach uses the inherent nonlinearity of measurement in quantum mechanics to generate entanglement. 
It is the latter approach that we will focus on in this review.
The rest of the paper is structured as follows. 
We first describe approaches based on solely using linear optics and measurement to generate entanglement. 
Sec.~2 describes relevant experimental components and considerations. 
Sec.~3 describes the theoretical and experimental progress in generating different classes of entangled states, along with methods for improving their efficiency. 
Sec.~4 details applications of heralded state generation. 
Conclusions are provided in Sec.~5.

\subsection{Heralding versus postselection}

\begin{figure*}
    \centering
    \includegraphics[width=\textwidth]{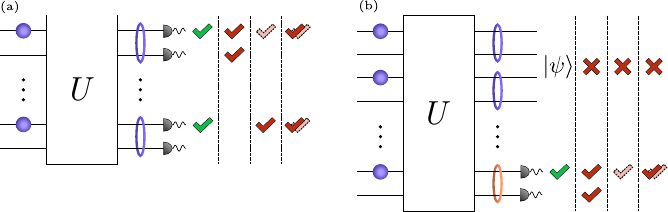}
    \caption{Postselected and heralded entangled state generation circuits.
    \textbf{(a)}: In postselected circuits, $n$ photons are injected into an $m$-mode linear optical circuit, $U$. Detectors are assigned to all modes such that all photons, notwithstanding loss, are measured. Typically, successful measurement patterns require only one photon being measured in certain groups of \textit{target} modes (circled in purple). This is shown after the detectors where ticks correspond to photon detections and dashed ticks correspond to undetected photons, which can be due to the loss of a photon or lack of photon number resolution in the detector used. Green denotes a successful outcome, and red denotes an unsuccessful outcome (which might be mistakenly categorised as successful due to losses or detector inefficiencies, as would be the case for the far-right scenario shown).
    \textbf{(b)}: In heralded entangled state generation, $n$ photons are injected into an $m$-mode circuit where a subset of modes, the \textit{heralding modes} (circled in orange), are measured. Through the measurement of a subset of photons acting as \textit{ancillas} in the heralding modes, the creation of an entangled state across the remaining, target modes is heralded. Since the photons encoding this state have not been measured, they can undergo further quantum information processing. A successful heralding pattern is shown verifying the creation of a state $\ket{\psi}$, whereas incorrect patterns denote a failed state generation attempt, denoted with a red cross in the target modes.}
    \label{fig:postAndHerEntanglement}
\end{figure*}

Unlike many other qubit platforms, measurement in photonics is generally a destructive process, achieved by the absorption of a photon by a material.
This means care is required when using photon measurement to generate entanglement. Broadly, the approaches can be classified into two categories illustrated in Fig.~\ref{fig:postAndHerEntanglement}.
In the first, all photons are measured, and successful state generation events are determined by the measurement pattern across all modes. 
As a result, the state is generated and consumed at the same time, and it cannot be used for further processing. 
This technique is commonly referred to in literature as \textit{postselected} entanglement generation.
The second category uses ancilla photons to indicate the circuit success without destroying the state. 
Measurement patterns are defined over a subset of the modes, and upon specific patterns being measured, the remaining photons 
 in the remaining modes form a state that can be used for further processing, unlike in postselected approaches. 
This technique is called \textit{heralded} entanglement generation and is the focus of this review. 

\subsubsection{Postselected entanglement generation}

We consider here postselected schemes of the form illustrated in Fig.~\ref{fig:postAndHerEntanglement}a, where all the input photons are measured following the linear optical circuit.
As a result, these schemes never actually output quantum states, but only measurement data that can be processed classically. 
\textit{Postselection} is the main processing used, where each measurement outcome is either kept unchanged or discarded depending on a postselection rule. 
Typically, the postselection rule rejects certain measurement patterns based on the numbers of photons detected in specific sets of modes. 
For instance, it is common to require each pair of modes used for dual-rail encoded qubits to contain exactly one photon to represent a valid state.  
In postselected schemes, all modes are referred to as target modes, since the state of interest (the target state) is identified within all modes.
By using a simple selection rule and separable input states, it is possible to generate the same statistics as those obtained with an entangled state, which is why this technique is commonly referred to as postselected entanglement generation. 
We will use this terminology in the rest of the review because it is commonly used in the literature, but we again highlight that there is no remaining entangled system at the end of a postselected procedure, since all photons have been detected and therefore destroyed. 

\begin{figure}[h]
    \centering
    \includegraphics[width=0.8\columnwidth]{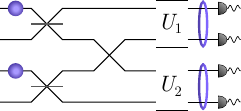}
    \caption{Postselected Bell state generator circuit. The beam splitters in each pair of modes are used to create a uniform superposition, then one mode from each dual-rail is swapped. $U_1$ and $U_2$ can be configured to measure the photons in a desired Pauli basis to reveal quantum correlations or characterise the state. The postselection rule only keeps measurement patterns where each pair of modes contains only one photon. }
    \label{fig:fourierInterferometerPSBG}
\end{figure}

The maximally entangled Bell states form an important class of two-qubit entangled states, which can be written as

\begin{equation}
    \ket{\Phi^{\pm}} = \frac{1}{\sqrt{2}} ( \ket{00} \pm \ket{11} ), 
    \label{eq:def_bell_state_phi}
\end{equation}
\begin{equation}
    \ket{\Psi^{\pm}} = \frac{1}{\sqrt{2}} ( \ket{01} \pm \ket{10} ).
    \label{eq:def_bell_state_psi}
\end{equation}
We can generate Bell states using passive optics and postselection.
One such scheme is illustrated in Fig.~\ref{fig:fourierInterferometerPSBG}. 
Here two single photons are injected into a 4-mode optical circuit that consists of two balanced beam splitters (BS), and a mode swapping of two modes, with one from each dual-rail qubit. 
Starting with the input state $\ket{\psi_\text{in}} = \ket{1010}_F$, the circuit and the postselection rule transform the state as follows: 
\begin{equation}
\begin{matrix}
    \ket{1010}_F \\
    \xrightarrow{\text{BS}}& \frac{1}{4}( \ket{1010}_F + \ket{1001}_F + \ket{0110}_F + \ket{0101}_F ) \\
    \xrightarrow{\text{Swap}}& \frac{1}{4}( \ket{1100}_F + \ket{1001}_F + \ket{0110}_F + \ket{0011}_F ) \\
    \xrightarrow{\text{PS}}& \frac{1}{\sqrt{2}}(\ket{1001}_F + \ket{0110}_F) \\
    &= \frac{1}{\sqrt{2}}(\ket{01} + \ket{10}). 
\end{matrix}
\end{equation}

It can be seen that after the postselection process, the $\ket{\psi^+}$ Bell state is obtained, as defined in Eq.~(\ref{eq:def_bell_state_psi}). 
Single-qubit unitaries $U_i$ can be introduced in each pair or modes to choose the bases the qubits are measured in, with the case of $U_i=I$ (identity) corresponding to a computational basis measurement. 
The observed statistics in different measurement bases can reveal strong correlations that are a signature of entangled states. The postselection rule is valid for any unitary choice, because the postselection rule is based on the overall number of photons within the dual-rail qubit, which is preserved by any single-qubit unitary. 
Hence, the postselection rule commutes with any single-qubit unitaries applied on the dual-rail qubits. 

The majority of experiments involving entanglement generation with photons to date have relied on postselection, sometimes implicitly~\cite{Schaeff2015,Wang2017,Bouwmeester1999,Pan2001,Zhao2004,Lu2007,Huang2011,Yao2012,Wang2016,Zhong2018,Erhard2018,Eibl2004,Mikami2005,Tashima2010,Kiesel2007,Prevedel2009,Wieczorek2009}. 
Typically, postselected schemes are experimentally simpler than heralded ones since fewer photons are required. 
However, there are two significant drawbacks. 
Firstly, since the state is consumed immediately as it is generated, it is not possible to retain it for further use. 
This prohibits the viability of multiplexing to increase the success rate, unlike in the heralded case, as we shall describe later. 

Secondly, postselection can make it impossible to concatenate steps that involve interference between different sets of modes (e.g., between different pairs of modes used for dual-rail encoding), such as is the case in entangling operations. 
This is because postselection requires checking that a specific number of photons is present in each set of modes, but the interference can modify these photon numbers. As a concrete consequence, the generation of certain entangled states, e.g.~certain graph states, is not possible using postselected entanglement, and in fact, the proportion of such states increases with increasing qubit number \cite{adcock_hard_2018}. 
Therefore, the use of postselection fundamentally limits accessible quantum states and the type of quantum information processing one can achieve.

\subsubsection{Heralded entanglement generation}

Heralded entanglement uses conditioning on the measurement outcome of a restricted set of modes, known as the heralding modes. 
These serve to indicate whether successful state generation has occurred or not.
The remaining output modes, the target modes, are not measured and support the target state, allowing further computation to proceed. 
We illustrate a generic scheme in Fig.~\ref{fig:postAndHerEntanglement}b.
The output of this procedure is an actual quantum state that can be used for further processing or stored in a quantum memory. 
We shall introduce specific schemes for heralded state generation in Sec.~\ref{sec:heralded_schemes}.
 
A concrete example was proposed in the work of Knill, Laflamme and Milburn \cite{knill_scheme_2001}. 
Here, nonlinearity is manufactured through a measurement operation on ancillary photons that interfere with signal photons,  creating a probabilistic interaction.
The gate introduced by KLM is the nonlinear sign (NS) gate, which provides a conditional sign change on the two-photon Fock component as 
\begin{equation}
    \alpha \ket{0}_{F} + \beta \ket{1}_{F} + \gamma \ket{2}_{F} \rightarrow    \alpha \ket{0}_{F} + \beta \ket{1}_{F} - \gamma \ket{2}_{F}.
\end{equation}
Note that for a purely linear optical transformation on one mode, the prefactor of the $\ket{2}_{F}$ term would be the square of $\ket{1}_{F}$ term. Two NS gates can be combined to create a CZ gate, which can create entanglement between two separable input photons with two ancilla photons. 

\subsubsection{Benefits of heralding}

Heralded schemes produce an independent signal that confirms the creation of the desired state. 
The heralding property enables mechanisms to detect some photon loss (see Sec.~\ref{subsect:imperfections}) and reject failed events, preventing their further processing. 
This verification is critical for schemes such as fusion-based quantum computation~\cite{bartolucci2023fusion}, where operations are conditioned on entanglement presence, directly supporting scalable and error-resilient computation (see Sec.~\ref{QC_subsect}). 
Metrological applications could also benefit from switching to remove incorrect photon states before they interact with a target sample (see Sec.~\ref{QuantMet}).

The generation of heralded entanglement is typically a probabilistic process, meaning it does not succeed on every attempt. 
This issue can be mitigated through an approach known as multiplexing. 
Here, separate attempts are combined to increase the overall success probability. 
By running multiple processes in parallel, either in space or time, one can pick the successfully generated state and route it to a predetermined set of modes (see Sec.~\ref{sub_sec:multiplexing}).  
With increasing resources, one can, in principle, get arbitrarily close to unit success probability. 
Multiplexing is crucial for creating large and complex states, such as graph states, required for fault-tolerant quantum computation and advanced algorithms~\cite{chen2024pub}. 
Moreover, heralded entangled states can be stored in quantum memories after successful generation, allowing deterministic usage at a later time \cite{Sangouard2011}. 
This property is critical for building large-scale quantum networks and quantum repeaters, enabling the storage and distribution of entangled states over long distances that would otherwise suffer from prohibitive decoherence and photon loss \cite{liu2021heralded}.

Entanglement swapping uses heralding to share entanglement between distant parties, which is a pivotal resource for quantum communications (see Sec.~\ref{QuantCom}).
This enhances the security of quantum communication protocols by confirming entanglement in a manner that reduces vulnerabilities to photon loss and eavesdropping, improving the fidelity of the shared states between parties.
Higher loss tolerances and fidelity are substantial in guaranteeing security and error correction across quantum communications applications, making heralded entanglement a cornerstone of scalable and robust communication protocols~\cite{han2022phase}.

\section{Experimental considerations}

\subsection{Photon sources}\label{subsec:photonsource}

\begin{figure}[h]
    \centering
    \includegraphics[width=\linewidth]{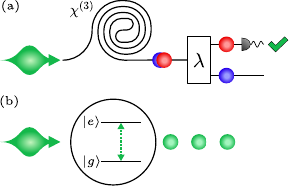}
    \caption{Single photon sources fall into two broad classes: pair sources (a) and single emitters (b). Pair sources emit photon pairs probabilistically through a nonlinear process such as spontaneous four-wave mixing, which uses a material's inherent optical nonlinearity in conjunction with an intense pump laser. Photons are separated using a degree of freedom of the photons, for example frequency using a spectral filter, and by measuring one photon in a pair, the presence of the other single photon can be heralded.
    Single emitters typically work by having an optical energy transition between some ground state, $\ket{g}$, and excited state, $\ket{e}$, whose energy difference corresponds to that of the desired single photon. As such, when the transition from the ground to excited state is pumped, for example optically with a laser, a single photon is emitted when the system relaxes from the excited to ground state.}
    \label{fig:sources}
\end{figure}

Generating constituent photons for entangled state creation can be achieved through different mechanisms and supporting hardware platforms. Photon sources generally fall into one of two broad classes: a) probabilistic sources of photon pairs, \textit{pair sources}, or, b) (in principle) deterministic single photons, typically called \textit{single emitters}.
Each class is depicted in Fig.~\ref{fig:sources}a,b respectively.

For the first class, nonlinear optical processes are employed in a variety of materials possessing a $\chi^{(2)}$ or $\chi^{(3)}$ component of their nonlinear susceptibility to generate photon pairs probabilistically.
These include spontaneous parametric down-conversion (SPDC) and spontaneous four-wave mixing (SFWM). 
SPDC is commonly used in bulk optics experiments with crystals such as potassium titanyl phosphate (KTP) \cite{Bayerbach_2023} or bismuth borate (BiBO) \cite{carolan_universal_2015}, and in integrated photonics through gallium arsenide \cite{Placke_AlGaAs_SPDC_2024} and lithium niobate \cite{Javid_LiN_SPDC_2021}.
SFWM has been used extensively in integrated photonics platforms and, unlike in $\chi^{(2)}$ processes, does not have the strict requirement of a non-centrosymmetric crystal structure. This means it can be observed in a wider range of materials, including silicon, silicon nitride, and indium phosphide \cite{Paesani_near_ideal_2020, Wu_SFWM_SiN_2021, Kumar_InP_SFWM_2019}, with silicon and silicon nitride being popular platforms due to their CMOS compatibility.
Compared to third-order nonlinearities, the magnitude of second-order nonlinearities tends to be higher, which results in these processes requiring less optical power for pair generation.
Optically pumping a nonlinear medium deterministically creates a single-mode squeezed vacuum (SMSV) or 2-mode squeezed vacuum (TMSV) state, depending on the nonlinear optical process taking place. The degree of squeezing is characterised by a squeezing parameter, $|\xi|$, which is proportional to the medium's relevant susceptibility component, interaction length, and power of the pump field(s).
Through expressing a TMSV state in the Fock basis,
\begin{equation}
    \ket{\xi} = \sech(|\xi|) \sum_k e^{ik\theta} \tanh^{k}(|\xi|)\ket{n}_{F,i} \ket{n}_{F,j},
\end{equation}
we observe a photon-pair summation, with the weight of each term depending on the squeezing parameter.
Here $\theta$ represents the phase of the pump field and $i,j$ denote the two modes \cite{Vogel_Welsch_Quantum_Optics}.
Whilst the generation of squeezed states in a nonlinear medium is deterministic, the generation of photon pairs is probabilistic. This is because the probability of generating one pair is determined by the amplitude of the one-pair term in the squeezed state.
The probability of obtaining one pair cannot be increased to unity by arbitrarily increasing the squeezing parameter, as for a high squeezing parameter, multi-pair terms become more probable.
As such, sources must be kept in the low-squeezing regime by keeping pump power sufficiently low to largely avoid generation events consisting of more than one pair of photons. This places an intrinsic limit on the number of single photon pairs that can be generated \cite{Schneeloch_2019}.
To knowingly obtain a single photon, one photon in the pair must be measured to herald the presence of the other photon in the pair, as depicted in Fig.~\ref{fig:sources}.

Examples of single emitters include: quantum dots \cite{Uppu_QD_source_2020}; colour centres such as nitrogen vacancy centers in diamond \cite{Smith_NV_source_2021}; and molecules \cite{Schofield_molecule_source_2022}.
These systems possess energy level transitions that can be optically driven to emit single photons, whose wavelength is determined by the energy level separation.
Emitters are typically placed inside cavity structures to enhance optical properties such as the proportion of emission into a particular radiative decay mode, as opposed to other unwanted radiative modes or non-radiative decay, enabling greater collection efficiency of photons.
In addition, the rate of emission is increased through Purcell enhancement caused by the presence of a cavity, reducing the radiative lifetime of an emitter \cite{Gritsch_23}.
Photons generated from single emitters can be directed with free space optics or coupled to fibres and integrated photonic circuits.
Some emitters can be created or embedded into materials where integrated photonic circuits are lithographically defined around them, enabling photons to be emitted directly into circuits \cite{Sartison_dot_integration_2022, Weng_NV_integration_2023}.

We now introduce several photon source characteristics that are key to heralded state generation.

\textit{Source efficiency}:
Heralding efficiency and brightness provide insight into the system efficiency of photon sources. These metrics can be related to state generation probability and the rate of false positive events. For pair sources, the heralding efficiency is defined as the probability of detecting a signal photon given its partner photon has been detected \cite{Klyshko_eff_1977,Lu_heralding_photons_Klyshko_2016}.
The heralding efficiency of a source is reduced by optical circuit loss. When detecting one photon for the purpose of using the other as a heralded single photon, the signal photon typically traverses a greater optical depth than the heralding photon, passing through more passive and or active optical components, and related losses should be minimized.
Integrated photonic chips can have superconducting detectors patterned onto them which negates the inevitable coupling losses experienced by photons being coupled off chip and sent to single-photon detectors \cite{Gyger2021}.
Single-photon detector efficiency also contributes to the heralding efficiency, effectively acting as an additional loss mechanism.
State-of-the-art pair sources have achieved heralding efficiencies in excess of 80\% \cite{Ramelow_herald_eff_13,slussarenko2017unconditional}.

For single emitters, brightness is typically defined as the product of the excitation, source and fibre-coupling efficiency, without taking into account detector efficiencies. This allows determining the fraction of photons that end up in a collection fibre given an excitation \cite{Tomm2021, Uppu_QD_source_2020, Margaria_pig_tailed_QD_2024}. The current record is 72\% in a semiconductor quantum dot \cite{Ding_brightest_source_2023}.
The brightness of a source can be enhanced by optimizing the excitation and collection efficiency by the use of an optical cavity, as described previously.

\textit{Indistinguishability \& Spectral Purity}: 

High-quality photonic quantum interference is vital to heralded entangled state generation schemes, with a direct impact on the fidelity of generated states.
Such interference requires photons to be indistinguishable in all degrees of freedom such as frequency, polarisation and time. 
The pairwise indistinguishability of photons originating from different sources, or from the same source at different times, influences the Hong-Ou-Mandel (HOM) interference visibility \cite{HOM_1987}.
Any mixedness in the state or distinguishability in any degree of freedom will result in a reduced visibility, whereas perfectly indistinguishable and pure photons will result in unit visibility \cite{Rarity_Tapster_SPDC_interference_1989, Grice_Walmsley_distinguishability_1997, Jones_distinguishability_2022}.
For single emitters, the highest reported (raw) HOM visibility is 98.5\% and was obtained with a quantum dot \cite{Ding_best_raw_HOM_2016}.

Pair sources possess a biphoton wavefunction that can be characterised.
This wavefunction is described by a joint spectral amplitude (JSA), which has a magnitude, $|\text{JSA}|$, and an associated phase, the joint spectral phase (JSP).
The JSP describes the phase relationship between the generated photons and both the $|\text{JSA}|$ and JSP can contain correlations that result in a degree of mixedness in the reduced state of a single photon from the pair. 
For the purposes of a heralded single photon source, an ideal JSA is uncorrelated in both its amplitude and phase, resulting in a pure single-photon state.
By performing a Schmidt decomposition, JSAs can be decomposed into a sum of discrete Schmidt modes \cite{Zielnicki_2018} with the Schmidt number, $K$, quantifying an effective number of modes. 
For an ideal source, $K=1$, and for a general source, the spectral purity is given by $P=1/K$.

The spectral purity of pair sources can be characterised by measuring the joint spectral intensity (JSI), defined as $\text{JSI}=|\text{JSA}|^2$  \cite{SET_Liscidini_2013}.
This provides a lower bound on the purity as it does not capture correlations that may exist in the JSP, though techniques do exist for JSP characterization \cite{Faruque2023,borghi_phase_resolved_2020}.
Another method to measure the spectral purity of photons from pair sources is to measure the unheralded second-order correlation function, $g_{u}^{(2)}$, which captures both amplitude and phase information of a JSA.
This is performed by blocking the path of one photon to effectively trace out one subsystem, leaving the other in a state whose statistics approach that of a thermal state for increasing purity.
For a given JSA, $g_{u}^{(2)}(0) = 1 + \frac{1}{K}$, at zero time delay, which for an ideally pure state with $K=1$ gives $g_{u}^{(2)}(0) = 2$, the same result as for a thermal state. 
Spectral purity is then given by $P = g_{u}^{(2)}(0) - 1$ \cite{Christ_2011}.
The highest reported spectral purity of a pair source measured via a JSI is 99.7\%, and in the same work, the HOM visibility between heralded photons from two different sources is 99.5\% \cite{PsiQ_manufacturable_platform_2024}.

Pair sources are typically used in two ways: either both photons generated in a spontaneous process are used in subsequent optical processing, or one is detected to give a heralded single photon.
If multiple pair sources are being used in the first manner and biphoton interference occurs, a high interference visibility requires a high degree of indistinguishability between the biphoton states from the two sources.
If multiple pair sources are used in the second manner described, the heralded single photons require a high degree of both indistinguishability and purity to achieve high HOM interference visibilities.
To measure the mutual indistinguishability of biphoton states from two different sources, reverse-HOM interference can be performed or the overlap of their JSIs can be used to give an upper-bound (given the lack of phase information in the JSI).
The indistinguishability of heralded single photons from different sources can be measured by performing HOM interference just as is done with single photons from one source \cite{Paesani_near_ideal_2020}.

\textit{Number Purity}:

For a given excitation, an ideal photon source should produce one photon. 
In practice, undesired multiphoton emissions can occur, introducing errors into heralded entangled state generation circuits.
For pair sources, the squeezed states generated in SFWM or SPDC in the Fock basis may have non-zero contributions from higher-order terms, corresponding to multiple pairs of photons being generated.
For single emitters, multiphoton emissions can occur due to an excitation pulse causing emission, re-excitation and a second emission, or due to phonon-assisted processes, for example \cite{Fischer2017}.
Additionally, for both classes, insufficient filtering of a pump laser via frequency, polarisation or some other degree of freedom can lead to pump light leaking into optical circuits, causing additional errors.

The number purity of a source is understood through its photon number statistics.
By performing a second-order correlation function, $g^{(2)}$, measurement, a true single photon source should have sub-Poissonian statistics, with a $g^{(2)}=0$ at zero time delay.
$g^{(2)}$ measurements can be performed using a Hanbury Brown Twiss (HBT) interferometer and performing coincidence measurements \cite{Schartz_g2_HBT_on_chip_2018}, where a $g^{(2)}>0$  indicates some degree of multiphoton contributions.
For pair sources, the $g^{(2)}$ measurement is conditioned upon measurement of the heralding photon, with the signal photon being sent into a HBT interferometer; this is usually referred to as a heralded $g^{(2)}$ measurement \cite{Christ_2011}.
The lowest reported $g^{(2)}$ for a single emitter is $g^{(2)} = 7.5\times10^{-5}$ \cite{Schweickert_best_dot_g2_2018} in a quantum dot, and for a pair source is $g^{(2)} = 9.4\times10^{-4}$ \cite{wang2024brightheraldedsourcereaching}.

\subsection{Photon detectors}

In heralded state generation, the heralding signal is based on a pattern of outcomes obtained from single-photon detectors that monitor the heralding modes.
State generation schemes are usually designed for ideal detection systems, which yield a heralding signal if and only if the appropriate numbers of photons impinge on the detectors.
Thus, the capabilities and properties of real detectors, which fall short of ideal behaviour, strongly influence the heralded state generation performance. 
This typically manifests in two ways: 1) missing real heralding events, thereby lowering the experimental success probability, and 2) the heralding of incorrect events, where the desired state is not present in the target modes.
Such errors are depicted in Fig.~\ref{fig:postAndHerEntanglement}(b).
The impact of detector imperfections on heralded state generation will be covered in Sec.~\ref{subsect:imperfections}, whilst here we provide an overview of different detector types and their capabilities.

We begin with a brief explanation of the terms we will use. 
The \textit{dark count rate} is the average rate of registered photon counts without any incident light. 
These false detection events typically originate from thermal and electrical noise. 
The \textit{dead time} is the time interval, after a detection event, that a detector spends in its recovery state before being able to measure another photon and produce a count. 
\textit{Timing jitter} is the uncertainty in the timing of recorded photon detection events. 
Zero timing jitter corresponds to an exact resolution of the arrival time of a photon, while a finite timing jitter leads to a spread of detection times. 

\subsubsection{Threshold and photon-number resolving detectors}
Photon detectors can broadly be divided into two categories, threshold detectors and photon-number-resolving (PNR) detectors.
Measurements with threshold detectors provide a binary outcome, which is usually the presence or absence of photons. 
PNR detectors, on the other hand, are capable of distinguishing and quantifying, up to a certain number, the number of photons measured at a given time.

\subsubsection{Pseudo-photon-number-resolving detection}
\begin{figure}[ht]
    \centering
    \includegraphics[width=0.6\columnwidth]{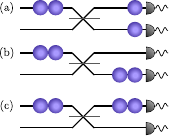}
    \caption{The simplest example of a pseudo-PNR detection set-up. (a) With a 50\% probability, the two incident photons are split by the beam splitter and are resolved and registered as a two-photon event. (b) and (c): With 25\% probability each, the two photons arrive together at one or the other detector and are thus not registered as a two-photon event.}
    \label{fig:fanout}
\end{figure}

Pseudo-PNR detection refers to devices or methods that have some ability to resolve photon numbers despite using threshold detectors. 
There are different techniques to realise pseudo-PNR detection; the main one involves fanning out one single optical mode into multiple modes, each being monitored by a threshold detector.
This method has limitations, namely: 1) It can only probabilistically register the correct number of incident photons because the number of incident photons per fanned-out mode can be larger than one. 2) It cannot register photon numbers greater than the number of modes fanned out into.
We show a simple example in Fig.~\ref{fig:fanout}; dividing one path mode into two by a balanced beam splitter followed by two threshold detectors is the most basic pseudo-PNR detection that can be realised. This scheme provides a 50\% probability of resolving two incident photons, with no ability to resolve more than two photons. 

\subsubsection{Superconducting nanowire single-photon detectors:}
 
Superconducting nanowire single-photon detectors (SNSPDs) include a superconducting nanowire that is cooled below its critical temperature, such that it enters a superconducting state and possesses zero electrical resistance.
The nanowire is biased with a DC current to just below its superconducting critical current, such that when a photon impinges on the nanowire, a localised non-superconducting hotspot with finite electrical resistance forms, generating a measurable voltage pulse.
These devices have evolved into a firmly established technology, demonstrating exceptional performance across an extensive range of wavelengths, enabling crucial experiments in various quantum optical applications.
SNSPDs can attain a system detection efficiency surpassing 99\%~\cite{chang2021detecting}, a dark count rate of approximately 0.25 counts per hour~\cite{Wollman20117}, a timing jitter below 3 ps~\cite{korzh2020demonstration}, and a recovery time of 500 ps.
However, the achievement of all of these numbers simultaneously on a single device is yet to be realised.
Typically, SNSPDs are used as threshold detectors. 
Nevertheless, it has recently been shown that using signal analysis of threshold SNSPDs, it is possible to extract some number resolution~\cite{Zhu2020Resolving}.
Moreover, pseudo-PNR can be achieved in special configurations, such as a parallel SNSPD architecture. 
For example, four photons have already been resolved in this manner~\cite{Stasi2023}. 

\subsubsection{Transition-edge sensors:}
The most famous example of PNR detection is the transition-edge sensor (TES). 
Superconducting TESs leverage the pronounced sensitivity of resistance to temperature changes in a superconducting material to quantify the energy of an incoming particle.
TESs, operating as calorimetric devices, grapple with a demanding trade-off between sensitivity and timing performance.
TESs can resolve up to 20 photons~\cite{li2023high} with a system detection efficiency of 95\%.
Some of the highest reported numbers include a timing jitter of 4 ns~\cite{Lamas-Linares2013}, recovery time near 1 $\mu$s~\cite{Calkins2011}, and quantum efficiency of $>98\%$~\cite{Lamas-Linares2013}. 

\subsubsection{Avalanche photodiodes:}
Despite their great efficiencies, SNSPDs and TESs both require cryogenic systems, running at temperatures below 5 K and 0.4 K, respectively~\cite{You2020,DeLucia2024}. 
Their efficiencies are also often polarisation-dependent.
Single-photon avalanche photodiodes (APDs) operate at room temperature (though high-efficiency telecom SPADs require some cooling).
They consist of reverse-biased p-n junctions set above the breakdown voltage to detect single photons.
They offer a wide photoactive area, are insensitive to the polarisation of light, and function in a wide range of wavelengths. 
Conversely, their efficiencies are lower than superconducting detectors.
They can reach an efficiency of approximately 60\%, dark count rates below 20 counts per second, and timing jitter of approximately 30 ps~\cite{ceccarelli2021recent}.

\subsection{Impact of imperfections}\label{subsect:imperfections}

\subsubsection{Loss and detector imperfections} 
Loss of photons is often the main source of noise in photonic schemes.
Scattering or absorption, imperfections in optical components, coupling light between systems and detectors, ineffective detection, and mode mismatches are just a few examples of how loss can arise \cite{SparrowPhD2018}.
Some mitigation for this can be made through refinement of component design and fabrication. 

Loss has a negative impact on many aspects of state generation, such as degradation or even complete loss of the attempted state.
Within heralded schemes, different effects can be seen, dependent on whether the loss occurs on the heralded modes or the target modes. 
Photons that are meant to be part of the target state are lost even though they were originally present, altering the heralded state (heralding of the wrong state) and reducing its fidelity with the target state.
We essentially get a \textit{false positive} result with the heralding signal suggesting we have created the state, which is not true in practice. 

When there is loss on the heralding modes, it is possible that a photon that would be required to herald the creation of the desired state is not detected, so we discard this state, thus reducing the success probability of the heralded generation. 
This is known as a false negative. 
On the other hand, loss on heralding modes can also produce a false positive, where a heralding signal is registered even though the incorrect state is generated. 
This can happen if there are more photons in a heralding mode than specified by the heralding pattern, but where the additional photons that would signify that the state has not been produced are lost. The correct heralding pattern is seen, even though the incorrect photonic state is present across the target modes.

Detector imperfections, such as dark counts, adversely affect the quality of the generated states. A number of heralded entangled state generation schemes utilise the heralding of zero photons, that is, heralding on vacuum.
Whilst this is conceptually simple, the experimental detection of zero photons poses several challenges.
In this scenario, because the absence of a detection is used for heralding, dark counts and inefficiency errors negatively impact the heralded output state.
The occurrence of a dark count will be recorded as a false negative, prompting the decision to discard the output, even if the right state was transmitted.
This diminishes the success probability of the device. 
In the case of heralding on photon presence in the heralding mode, dark counts can cause a false positive on the heralding mode, where there is no real photon, thus flagging the incorrect state in the target modes, and reducing the fidelity of the output state. Fidelity and success probability are both degraded through 
dark counts. 
Increasing dark counts leak into the prescribed detection window, causing false positives (negatives) depending on whether heralding occurs on a photon (the vacuum). 
This can be mitigated by gating the detection time window through an external clock, such as a laser pulse.

Other forms of detector imperfections, such as long jitter time and dead time, have negative effects on the quality of the generated state. 
Large dead times reduce the maximum entangled state generation rate if they are longer than the laser repetition period (assuming a deterministic photon source). 
Longer timing jitter of the detector increases the uncertainty in the timing of the heralding photon(s). 
This results in the reduction or loss of temporal correlation between the heralding signal and the target states.

\subsubsection{Distinguishability} 
 
We introduced photonic indistinguishability within the context of photon sources in Sec.~\ref{subsec:photonsource}. 
In practice, no two photons will be perfectly indistinguishable and no circuit will operate completely without error. 
Irrespective of whether photon distinguishability is introduced at the single photon generation level or later in the computation, correlations are produced between the degree of freedom in which the photon is computationally encoded and other, internal degrees of freedom. 
This leads to the degradation of desired interference required for heralded entangled state generation, for example through the introduction of which-path information \cite{Mandel:91, Jones_distinguishability_2022}. 
This information can be linked to the HOM visibility, providing us with a tool to realistically model internal photonic state structures. 
Two models that were introduced by Sparrow are the \textit{orthogonal bad-bits} (OBB) and \textit{random source} (RS) models \cite{SparrowPhD2018}. 
These models can provide a description of imperfections in photon sources arising from distinguishability. 
Such cases include spectrally mixed photons from spontaneous sources and single photons at differing emission frequencies from separate deterministic emitters. 
More recently, several adaptations of these models, which place further constraints on the internal states provided by RS and OBB for different scenarios, have also been introduced \cite{saied2024general, Saied_2024}.

Whilst this review focuses on heralded entangled state generation, it is important to highlight the impact of distinguishability errors on postselected states. 
This will provide further weight to the subtleties between postselected and heralded entanglement generation. 
In \cite{SparrowPhD2018}, it is shown that for the postselected Bell state generation scheme introduced in Fig.~\ref{fig:fourierInterferometerPSBG}, the impact of partial distinguishability is to apply an effective dephasing channel to one qubit of the desired two-qubit Bell state. 
Vitally, this error mechanism is derived from restricting the output state and the effect of partial distinguishability to the two-qubit computational subspace, spanned by the tensor product of single-qubit logical states introduced in Eq.~(\ref{eq:def_dual_rail}) as $\{\ket{00},\ket{01},\ket{10},\ket{11}\}$. 

When moving to heralded linear optical circuits, we can no longer simply omit state contributions from non-computational terms where more than one photon occupies the same dual-rail encoded qubit modes. 
Such mode occupation is known as \textit{non-computational leakage}, leading to further state degradation relative to that considered in the postselected regime \cite{shaw_errors_2023}. 
These errors may occur even when the heralding signal suggests the desired state has been produced. 
One proposal to tackle leakage errors in the context of photonic quantum computation is through the application of $n$-GHZ state analyzers introduced in \cite{bartolucci_creation_2021, gimeno2016towards}. 
Whilst such techniques are broadly effective, at the \textit{resource state} generation level where we generate of the order 10-qubit 
entangled states (see Sec.~\ref{subsubsec:fbqc} and \cite{bartolucci2023fusion}) from Bell and GHZ \textit{seed states}, leakage errors may contaminate gate operations and corrupt subsequent computation. 
Although there are now multiple works, both analytic and numerical, modelling distinguishability errors in the context of photonic quantum information processing \cite{rohde_error_2006, SparrowPhD2018, shaw_errors_2023, saied2024general, Saied_2024}, these are restricted to small numbers of qubits and computational steps. 
The complexity in faithfully representing photonic distinguishability is a main contributor to this, with future work required in this area to ameliorate simulation overheads.

\subsubsection{Multiphoton errors}
Both pair sources and single emitters can spuriously emit more than one (heralded) photon.
As discussed in Section \ref{subsec:photonsource}, when pair sources are pumped a squeezed state is created which in the Fock basis is composed of terms with an in integer number of photon pairs. 
Even when pumping in the low-squeezing regime, there remains a finite probability of greater than one pair of photons being emitted in a nonlinear process.
This results in multiple photons occupying the same optical mode entering an optical circuit unless photon number resolution is available when the herald photons are measured.
Even then, if loss occurs on a subset of the herald photons, this could produce a false-positive heralding pattern when in fact more than one signal photon has entered an optical circuit.
For single emitters, multiple photons can be emitted for different reasons. For example, if an emitter has a short lifetime compared to the pump pulse that is exciting it, a single pump pulse could cause multiple excitation and emission events as it interacts with the emitter.
Unlike pair sources there is not a straightforward heralding signal that can be used to detect such errors from occurring.

These additional photons generated by sources increase the noise in an optical system and can reduce the fidelity of a target state or could even end up in the target modes to produce the incorrect state.

\section{Schemes and implementations for heralded state generation}
\label{sec:heralded_schemes}

So far, we have provided a brief introduction to the framework of heralded state generation. 
We now turn to the generation of specific heralded entangled states.
Different circuits and schemes have been proposed to generate entangled states. This section considers the experimental and theoretical schemes to generate these states, as well as their related success probabilities. 
We then discuss ways to improve these success probabilities, along with methods to find these circuits.

\subsection{Bell states} \label{sect:BSG}

The four Bell states, given in Eqs.~(\ref{eq:def_bell_state_phi}) and (\ref{eq:def_bell_state_psi}), form a maximally entangled basis for two qubits. Bell states can be used in numerous quantum information protocols, such as in teleportation, entanglement swapping and as a resource state for linear optical quantum computing. 

\subsubsection{Bell state generation schemes}

The first schemes for heralded generation of Bell states were realised in bulk optics using photons produced from SPDC sources \cite{Sfiliwa2003Conditional, barz_heralded_2010, wagenknecht_experimental_2010}.
An optical circuit in conjunction with a triple photon pair emission from an SPDC source are used to herald the creation of the $\ket{\Phi^+}$ Bell state, with entanglement in the polarisation basis.
Detection of one photon in each of the four heralding modes, making up a four-fold coincidence, heralds the generation of the state.
The creation of the states was experimentally verified using quantum state tomography measurements, allowing determination of the associated state fidelity. 
In the set-up used by both groups, they aim to limit the effect of multi-photon errors by tuning the beam splitter reflectivities to improve their success probabilities. 
The experiment carried out in \cite{barz_heralded_2010} obtains a maximum fidelity of $F = (84.2 \pm 8.5)\%$, whereas the experiment carried out in \cite{wagenknecht_experimental_2010} obtains a maximum fidelity of $F = (88.2 \pm 2.8)\%$. 

Following these demonstrations, and with the development of integrated on-chip optical technologies, a number of optical schemes using different photon generation processes, such as SFWM, have been developed.

\begin{figure}
    \centering
    \includegraphics[width=\columnwidth]{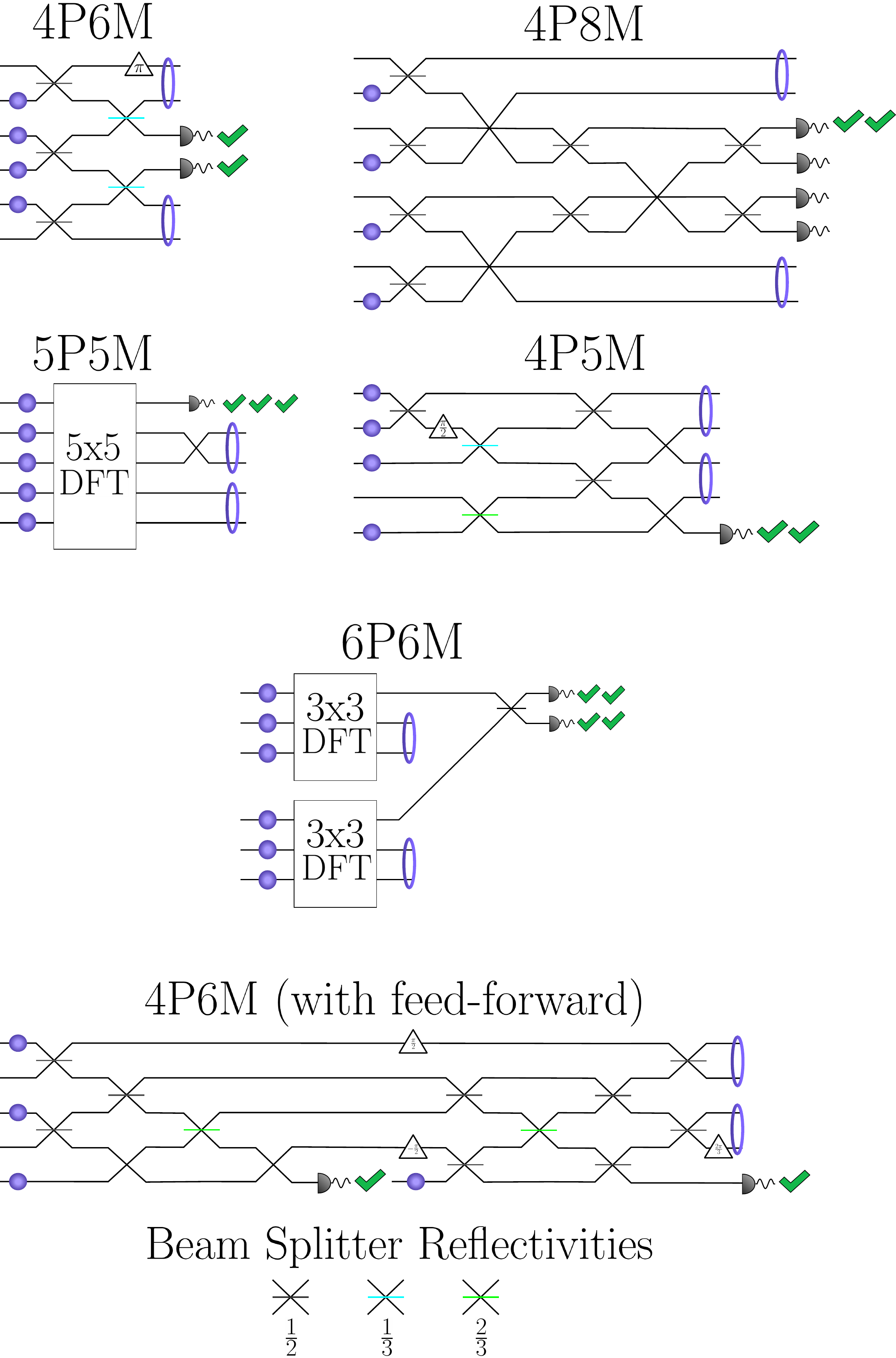}
    \caption{Schemes for heralded Bell state generation. Ticks suggest the (or, in the case of multiple options, one) detector ``click'' pattern that heralds the generation of a Bell state on the target modes. The labels $x$P$y$M indicate that the schemes are using $x$ photons in a linear optical circuit comprising $y$ modes, and DFT represents a linear optical circuit to carry out the discrete Fourier transform. } 
    \label{fig:bsg_schemes}
\end{figure}

In the scheme presented by Carolan \textit{et al.}~\cite{carolan_universal_2015}, a heralded Bell state generator using four photons in six modes (4P6M) is implemented, using the linear optical circuit shown in Fig.~\ref{fig:bsg_schemes}.
Detection of photons in both of the central modes heralds the creation of the $\ket{\Phi^+}$ state across the remaining modes, with a success probability of $p_{\text{success}} = 2/27$.
This scheme was implemented on an integrated photonic chip with an off-chip photon source, demonstrating the generation of the Bell state through tomography.

A scheme using four photons in eight modes (4P8M) is proposed in \cite{Zhang2008Demonstration, bartolucci_creation_2021}.
This scheme, shown in Fig.~\ref{fig:bsg_schemes}, generates the $\ket{\Phi^+}$ state, heralded by the detection of single photons on any two out of the four heralding modes, with vacuum on the remaining two heralding modes, leading to six possible heralding patterns.
This results in a success probability of $p_{\text{success}} = 3/16$; however, this can be improved through the application of techniques such as boosting and bleeding, as described in Sec.~\ref{sub_sec:multiplexing}.

A more general scheme for creating qudit entangled states is given by Paesani \textit{et al.}~\cite{paesani_scheme_2021}.
Five input photons in five modes (5P5M) can be used to generate a Bell state.
This scheme is comprised of a 5$\times$5 discrete Fourier transform (DFT) linear optical circuit.
Creation of the $\ket{\Psi^+}$ state is heralded by three photons being detected in the one heralding mode.
Unlike the 4P8M or 4P6M schemes, this scheme requires a PNR detector on the heralding mode.
The scheme performs with a success probability of $p_{\text{success}} = 12/125$. 

DFTs are also used in the scheme laid out by Bhatti and Barz in \cite{bhatti_heralding_2024}. 
Here, they propose multiple Bell state generator schemes based on N-mode symmetric multi-port beam splitters (SMSs). 
These SMSs implement a unitary transformation, the elements of which are defined using a DFT. 
To generate a Bell state, a 6-mode interferometer is used with a single photon on each input mode as shown in Fig.~\ref{fig:bsg_schemes}. 
These are then incident on two 3-mode SMSs and the top mode of each SMS is routed to a final 50:50 beam splitter. 
If four photons in any pattern are detected by the two PNR detectors on the outputs of the beam splitter, this heralds the creation of a  $\ket{\Phi}$ state with $p_{\text{success}} = 4/27$. Whether the $\ket{\Phi+}$ or $\ket{\Phi-}$ is created depends on the heralding pattern.
A Bell state generator scheme using four input photons and 4-mode SMS circuits is also proposed in \cite{bhatti_heralding_2024}. Although this only requires threshold detectors, its success probability is lower, with $p_{\text{success}} = 1/8$, and it requires input photons in two different internal degrees of freedom.

SMS linear optical circuits, using a similar structure to that shown in Fig.~\ref{fig:bsg_schemes}, can also be used to generate higher-dimensional Bell states. 
For $d$ dimensions, $3d$ input photons are required. 
One photon is incident on each mode of $d$ 3-mode SMSs. 
The top mode of each 3-mode SMS is then routed to a $d$-mode SMS. The outputs of the  $d$-mode SMS are then incident on PNR detectors. 
If $3d-2$ photons are detected using the PNR detectors, this heralds the creation of a $d$-dimension Bell state across the remaining modes of the $d$ 3-mode SMS schemes. 
The success probability is given by $p_{\text{success}} = \frac{d \times 2^{d-1}}{3^{2d-1}}$ \cite{bhatti_heralding_2024}. 
The bleeding method laid out in Sec.~\ref{subsubsect:bleeding} can be applied to this scheme, in order to increase the success probability to $p_{\text{success}} = \frac{d \times \left(2 + 2^{d-1} \right)}{3^{d}}$.
Fldzhyan \textit{et al.}~\cite{fldzhyan_compact_2021} present a more compact scheme, using four photons in five modes (4P5M), with an optimization process to minimise the number of optical components in the circuit. 
This scheme, as shown in Fig.~\ref{fig:bsg_schemes}, generates the $\ket{\Phi^+}$ state, conditional on heralding two photons in the single heralding mode.
Similarly to the 5P5M scheme, this scheme requires PNR detection.
The scheme has $p_{\text{success}} = 1/9$.

Each of the aforementioned schemes requires the photons to be fed at once to the circuit input. 
Schemes also exist where a subset of input photons are fed into the circuit at a later time, conditioned on the observation of intermediate heralding events.
For instance, such a feed-forward scheme has been proposed in \cite{fldzhyan_compact_2021} using four photons in six modes, as illustrated in Fig.~\ref{fig:bsg_schemes}.
Initially, three photons are input to the first section of the circuit. 
If one photon is measured on the heralding mode of the first section, the procedure moves on to the second section of the circuit, injecting an additional fresh photon.
Detecting a single photon again on the heralding mode of the second section heralds the generation of the $\ket{\Phi^+}$ state, with a success probability of $p_{\text{success}} = 2/27$.

We summarise the success probabilities associated with the schemes described above in Table \ref{table:bsg_schemes}. Techniques to improve these numbers have been proposed and are described later in Sec.~\ref{sub_sec:multiplexing}. 
However, these techniques often present experimental challenges, including PNR detection, the coherent control of many photons, circuitry and switching.

\begin{table*}[htb]
\begin{tabularx}{\textwidth}{*{6}{| >{ \raggedright\arraybackslash}X}{|>{\raggedright\arraybackslash}X |}}
   \toprule
Scheme reference & Number of input photons $N$ & Number of modes & Success probability & Detector type\\ 
   \midrule
\cite{barz_heralded_2010} & 6 & 8 & Varies with beam splitter transmission & Threshold \\
\hline
\cite{wagenknecht_experimental_2010} & 6 & 14 & Varies with beam splitter transmission & Threshold \\
\hline 
\cite{carolan_universal_2015} & 4 & 6 & $2/27 \approx  7.40\% $  & Threshold\\
\hline
\cite{bartolucci_creation_2021} & 4 & 8 & $3/16 \approx  18.7\% $  & Threshold\\
\hline
\cite{paesani_scheme_2021} & 5 & 5 & $12/125 \approx  9.60\% $  & PNR\\
\hline
\cite{bhatti_heralding_2024} & 6 & 6 & $4/27 = 14.8\% $  & PNR \\
\hline
\cite{fldzhyan_compact_2021} & 4 & 5 &  $1/9 \approx 11.1\%$ & PNR \\
\hline
\cite{fldzhyan_compact_2021} (w/ feed-forward) & 4 & 6 & $2/27 \approx 7.4\% $ & Threshold \\
   \bottomrule
\end{tabularx}
\caption{Comparison of existing schemes (as shown in Fig.~\ref{fig:bsg_schemes}) for heralded Bell state generation.}
\label{table:bsg_schemes}
\end{table*}

\subsection{Schemes and implementations for heralded NOON state generation}

NOON states are entangled 2-mode states with a definite photon number, $N$, where one has a balanced superposition of all photons being in one mode or the other. 
In the Fock basis, they are defined as 
 \begin{equation}
     \ket{\mathrm{N00N}(N)} = \frac{1}{\sqrt{2}} (\ket{N0}_{F} + \ket{0N}_{F}).
     \label{eq:def_noon_state}
 \end{equation}
 
NOON states possess maximal variance of the photon number difference between the two modes, $\hat{n}_a-\hat{n}_b $; as a consequence, they are optimal probes for phase estimation tasks.  
In the textbook scenario where there is no photon loss, NOON states reach the Heisenberg limit, the ultimate precision achievable in quantum metrology. 
More intuitively, this enhanced phase sensitivity arises from the fact that each creation operator in the state picks up the phase that would be imparted on a single photon.

\subsubsection{NOON state generation schemes}

The special case of the $N=2$ NOON state can be created deterministically from two single photons via HOM interference at a 50:50 beam splitter, which results in two-photon bunching. By contrast, NOON states with larger $N$ cannot be prepared deterministically from separable inputs \cite{Parellada2023}; this implies that at least three modes are required within a heralded approach. 
A flurry of activity in the early 2000s led to the discovery of many different schemes for the preparation of heralded NOON states with $N>2$, with a summary of key properties provided in Table \ref{table:noon_schemes}.

\begin{table*}[htb]
\begin{tabularx}{\textwidth}{*{6}{| >{ \raggedright\arraybackslash}X}{|>{\raggedright\arraybackslash}X |}}
   \toprule
Scheme reference & Number of photons in NOON state, $N$ & Number of input photons & Number of modes, where $n \in \mathbb{Z}$ & Success probability	& Vacuum heralding requirement & Detector type\\ 
   \midrule
\cite{Fiurasek2002}   & $N \in \mathbb{Z}$ & $N$ & $N+2$ for $N = 2n$ or $2N + 2$ for $N = 2n + 1$ &  $\frac{3}{16}= 18.75\%$ for $N = 4 $ & Yes & Threshold \\
\hline
\cite{Lee2012}  & $N = 2^{k}, k \in \mathbb{Z}$ & $N$ & $4$ for $N = 4$ & $\frac{3}{16}= 18.75\%$ for $N = 4 $ & Yes & Threshold \\
\hline
\cite{Pryde2003} & $N \in \mathbb{Z}$ & $N$ & $N$ & Not provided & Yes & Threshold \\
\hline
\cite{Hofmann2004}   & $N \in \mathbb{Z}$ & $N$ & $2N$ & $\frac{3}{256}\approx 1.17\%$ for $N = 4 $ & Yes & Threshold \\
\hline
\cite{vanmeter_general_2007}  & $N = 5$ & $6$ & $4$ & 5.64\% for $N=5$ & Yes & PNR \\
\hline
\cite{Lee2002Nupto4}  & $N \in [2,3,4]$ & 6  & $4$  & $\frac{3}{64}\approx 4.69\%$ for $N=4$ & No & PNR \\
\hline
\cite{Zou2002}  & $N = 4k, k \in \mathbb{Z}$ & $\frac{3N}{2}$ & $\frac{N}{2} + 2$ & Not provided & No & PNR \\
\hline
\cite{Kok2002}  & $N = 4k, k \in \mathbb{Z}$ & $2N$ & $N+2$ for $N = 2n$ or $2N + 2$ for $N = 2n + 1$ & Not provided & No for even $N$, yes for odd $N$ & PNR\\
   \bottomrule
\end{tabularx}
\caption{Comparison of existing schemes for heralded NOON state generation.}
\label{table:noon_schemes}
\end{table*}

\emph{Schemes with $N$ input photons}---Interestingly, the generation of NOON states allows for schemes that require only $N$ photons as input, where success is conditioned on obtaining only vacuum in the heralding modes \cite{Fiurasek2002,Pryde2003,Hofmann2004,Lee2012}. 
Each of these schemes uses an input of $N$ single photons distributed across $N$ modes. 
The schemes of \cite{Fiurasek2002,Pryde2003,Hofmann2004} apply to the creation of NOON states with arbitrary $N$, whereas that of \cite{Lee2012}  directly applies only to $N=2^k$, $k\in \mathbb{N}$ (other values of $N$ can be achieved but require more than $N$ input photons). 
The schemes differ in the total number of modes involved (and consequently also in the number of heralding modes), as well as in the associated success probabilities. 
This information is provided in Table \ref{table:noon_schemes}. 
As a tangible example, for the case of $N=4$, the success probabilities are 3/16 for the schemes in \cite{Fiurasek2002,Lee2012}, and 3/256 for the scheme in \cite{Hofmann2004}, while the number of total modes is four for \cite{Lee2012}, six for \cite{Fiurasek2002}, and eight for \cite{Hofmann2004}.

\emph{Schemes with $N+1$ input photons}---VanMeter \textit{et al.}~identified an example of heralded NOON state generation using an input of $N+1$ photons, namely the generation of the $N=5$ NOON state from an input of $\ket{2,2,2,0}_{F}$ \cite{vanmeter_general_2007}. 
This scheme relies on heralding $[1,0]$ and so maintains in part the difficulties associated with vacuum heralding. 
However, the addition of one input photon compared to the pure vacuum heralding schemes of \cite{Fiurasek2002,Pryde2003,Hofmann2004,Lee2012}, brings some advantages to this scheme: it requires fewer total modes, 4, and offers a significantly higher success probability, 0.05639.

\emph{Schemes with $N+2$ input photons}---Lee \textit{et al.}~\cite{Lee2002Nupto4} provided further schemes that apply to specific photon number choices $(2,3,4)$, rather than arbitrary values of $N$. 
For $N=4$, one of their proposals uses an input of $\ket{3,3,0,0}_{F}$ whilst another uses $\ket{2,2,1,1}_{F}$, with both schemes relying on a heralding pattern of $[1,1]$. 
By using more input photons than \cite{Fiurasek2002,Pryde2003,Hofmann2004,Lee2012}, these schemes eliminate the need for any vacuum heralding. 
Despite these additional input photons, they avoid increasing the total number of modes, thanks to the multi-photon inputs into some of the modes. 
However, the success probability for $N=4$ drops from 3/16 in \cite{Fiurasek2002,Lee2012} to 3/64. 

\emph{Schemes with $\frac{3N}{2}$ input photons}---Zou \textit{et al.}~\cite{Zou2002} propose a set of closely related schemes for arbitrary $N$, with slight differences depending on the choice of $N$. 
For $N$ divisible by 4, they require $\frac{3}{2}N$ photons in, distributed as $\ket{\frac{N}{2},\frac{N}{2},1,1,\ldots,1}_{F}$, and the heralding is performed based on $\frac{N}{2}$-fold single-photon detections [1,1,\ldots,1], thus also avoiding vacuum heralding. A further advantage of this scheme is the low total number of modes involved, $\frac{N}{2}+2$. 
The authors also report a high success probability, $0.097$ for $N=6$, which exceeds that of the schemes in \cite{Fiurasek2002,Pryde2003,Hofmann2004}.

\emph{Schemes with $2N$ input photons}---On the higher end of input photon and mode requirements, Kok and colleagues proposed a scheme requiring an input of $2N$ photons that occupy only two modes, that is,  $\ket{N,N,0\ldots,0}_{F}$ \cite{Kok2002}. 
When $N$ is even, the scheme involves a total of $N+2$ modes. 
The heralding pattern consists of single photons over $N$ modes, thus avoiding vacuum heralding.

\subsubsection{NOON state generation experiments}

There have also been a number of experiments on the generation of NOON states.

\emph{Experiments with $N=2$}---The fact that two-photon NOON states can be obtained deterministically from two single photons creates the impression that heralding is unnecessary for $N=2$. 
However, this assumes ideal, truly deterministic single-photon sources. These are not generally available, even today, despite the rapid developments of quantum dots, as they require perfect single-mode coupling into specified modes with zero loss. Past experiments used SPDC as an alternative approach for NOON state generation.
The most straightforward approach for a heralded two-photon NOON state from SPDC photons is as follows: One produces the two single photons by detecting one photon from each of two SPDC sources, and lets the photons undergo HOM interference at a beam splitter. 
Smith \textit{et al.}~reported a bulk-optical implementation of this idea in \cite{Smith2008}. 

A different method for generating two-photon NOON states from four SPDC-produced photons was demonstrated by Eisenberg \textit{et al.}~\cite{Eisenberg2005}. 
This bulk-optics experiment made use of double-pair emission from an SPDC source configured as a non-heralded polarisation-based Bell state source. 
Similarly to the experiment of Smith \textit{et al.}~\cite{Smith2008}, the experiment involved four modes in total, with threshold detection sufficing. This is because, in the implemented scheme, heralding relies on one photon incident on each detector and multiple photons are not expected to arrive, except those due to higher-order SPDC terms.

Other experiments demonstrated the heralded preparation of families of states that contain the two-photon NOON state as a member. 
A bulk-optical experiment by Ra and colleagues \cite{Ra2015} implemented a scheme that can prepare states related to the NOON state by tuning the phase between, and amplitudes of, the NOON state terms. 
An interesting property of this approach is that the amplitude and phase controls are effectively integrated into the heralding mechanism. 
This property arises from the structure of the overall linear optical unitary, where the phase and amplitude controls are situated at the end and in the heralding modes, without subsequent interaction between heralding and target modes. 
Since the heralding modes can be spatially separated from the target modes, this, in principle, allows remote setting of the amplitude and phase. 
In their experimental implementation, Ra \textit{et al.}~demonstrated phase control for $N=2$. 
The experiment involved four modes, with the input defined as the product state $|2,2,0,0\rangle_F$, and a heralding pattern of $[1,1]$.

Similarly, Vergyris \textit{et al.}~demonstrated the heralded generation of a family of states containing the two-photon NOON state, using four input photons \cite{Vergyris2016}. 
This was an integrated optical experiment on a hybrid chip. 
It consisted of two heralded single-photon sources followed by a tunable beam splitter implemented as a Mach-Zehnder linear optical circuit with a tunable phase, which could interpolate between separable outputs and NOON states.

\emph{Experiments with $N=3$}---Mitchell \textit{et al.}~\cite{Mitchell2004} performed an early experiment in bulk-optics, implementing a scheme for vacuum-heralded three-photon NOON state generation. 
While the scheme in principle requires three deterministic input photons, the experiment used two photons produced by SPDC and another photon from a weak coherent state. This meant the input to the heralded scheme was only probabilistically available at unknown times. 
In addition, vacuum heralding with imperfect detectors is prone to producing false heralding signals. 
The authors navigated both of these challenges by postselecting, based on detecting three photons within the target modes. 
Thus, although the state generation scheme is a heralded protocol, photon-counting postselection was employed to ensure that the input state was available.  
The below NOON state experiments all share this need for photon-counting postselection. 
This is unlike the demonstrations of heralded two-photon NOON states \cite{Smith2008,Eisenberg2005}, where the heralding itself ensured that the right number of photons existed at some point in the experiment (up to higher-order terms). 

Another demonstration of three-photon NOON state generation was published in 2009 \cite{Kim2009}. 
This 3-mode bulk-optics experiment by Kim \textit{et al.}~relied on double pair emission from one SPDC source as the four-photon input, and was based on photon subtraction. 
Photon subtraction could in principle be heralded by detecting one photon in one heralding mode, thus avoiding the need for vacuum heralding. 
However, the experiment used threshold detectors and additionally, the required double-pair emission to produce the necessary input could only be verified in postselection, based on three photons arriving in the target modes. 

\emph{Experiments with $N=4$}---A four-photon NOON state was demonstrated by Matthews \textit{et al.}~\cite{Matthews2011}. 
Their experiment used six photons generated by bulk-optical SPDC, coupled onto a reconfigurable silica-on-silicon waveguide circuit. 
The detection of exactly one photon in each of two heralding modes, in conjunction with postselection of four photons in the two target modes, attested that six photons had been generated in the first place (up to higher-order events), and that the heralded scheme had operated successfully, resulting in a four-photon NOON state. 

The sensitivity of NOON states to loss limits their practical applications, motivating a search for states that are more robust to loss~\cite{Barbieri2022}.  

\subsection{GHZ states}
\label{subsec:GHZstates}

The Greenberger-Horne Zeilinger (GHZ) states are a family of entangled quantum states comprised of three or more particles. The simplest and most well-known example of a GHZ state is the three-qubit state 
\begin{equation} \label{GHZmainEq}
\ket{\mathrm{GHZ}} = \frac{1}{\sqrt{2}} (\ket{000} + \ket{111}).
\end{equation}
This can be extended to GHZ states with $N > 3$ qubits as $ \frac{1}{\sqrt{2}} \left(|0\rangle^{\otimes N} + |1\rangle^{\otimes N}\right)$, and further generalised from qubits to $d$-dimensional subsystems through

\begin{equation} \label{GHZhighdimEq}
   \ket{\mathrm{GHZ}}_{N,d}=\frac{1}{\sqrt{d}} \sum_{i=0}^{d-1} \ket{i}^{\otimes N}  .
\end{equation}

GHZ states can exhibit starker nonlocal behaviour than Bell states. This is because for GHZ states, quantum mechanics provides a deviation from local realism that can be seen with single-shot events without the need to repeat the experiment many times to evaluate statistical inequalities \cite{Greenberger1989,Greenberger1990}.
As the number of particles or subsystem dimensions increases, GHZ states allow a wealth of contradictions with respect to classical expectations \cite{Ryu2014,Lawrence2014}, for example based on symmetry arguments. 

Adding to this interest from a quantum foundations perspective, there is also a strong drive to create GHZ states---particularly heralded---for technological reasons. 
Most pertinently, heralded three-photon GHZ states can be used as a building block to create large multi-photon entangled resources without feed-forward requirements \cite{gimeno2015three}.
These resources, in turn, enable universal quantum computing via linear-optical quantum computers (LOQC) in the measurement-based paradigm (more details are provided in Sec.~\ref{QC_subsect}). 

\subsubsection{GHZ state generation schemes}

Given deterministic single-photon sources, a heralded three-photon GHZ state can be generated from a product state of six input photons.
In \cite{Varnava2008}, Varnava, Browne, and Rudolph proposed a method involving 12 optical modes, which has a success probability of $1/64$, producing the GHZ state of Eq.~(\ref{GHZmainEq}), with a $1/64$ probability to produce the related state $\frac{1}{\sqrt{2}} (\ket{000} - \ket{111})$, conditioned on a different heralding pattern set.
These heralding patterns involve detecting three single photons in three of six heralding modes, with vacuum in the remaining three heralding modes.
With feed-forward, the GHZ-like state $\frac{1}{\sqrt{2}} (\ket{000} - \ket{111})$ could be locally corrected and then combined with the other heralding events to produce the GHZ state of Eq.~(\ref{GHZmainEq}) with a success probability of $1/32$.
In \cite{gubarev_improved_2020}, Gubarev proposed an alternative scheme that also uses six single input photons, only requiring ten modes. The successful generation of a GHZ state has a probability of $1/54$ and is conditioned on patterns containing three single photons and vacuum across the four heralding modes. 

With probabilistic pair sources, in principle, one could start with 12 photons (six pairs), herald six single photons by detecting their respective partner photons, and use the above schemes.
However, other proposals exist for probabilistic photon-pair sources that improve on this naive approach.
An early scheme by Walther \textit{et al}.~\cite{Walther2007} requires 12 photons produced by entangled photon-pair sources and uses 24 modes, but importantly, threshold detectors with imperfect efficiencies suffice.
This scheme is also extendable to higher-photon-number GHZ states, with $2N$ photon-pair emissions (that is, $4N$ photons) required for an $N$-photon GHZ state.
An alternative scheme was proposed by Niu \cite{Niu2009}, which requires the more difficult ability to distinguish between 0, 1, or $>1$ photon(s) incident on detectors. The advantage here is that only 10 photons are required.
This scheme has a success probability of 1/16 if feed-forward and local unitary operations are available, and is also extendable to $N$-photon GHZ states, where $2N-1$ pair sources ($4N-2$ photons) are required.
Krenn \emph{et al.}~discovered that it is possible to produce three-photon GHZ states with 10 photons and 13 modes, without reliance on vacuum detection or photon-number resolution \cite{Krenn2021}.

General, including high-dimensional, GHZ states can also be produced using heralded generation schemes designed for deterministic photon sources, such as those presented by Paesani \textit{et al.}, Chin \textit{et al.}~and Bhatti and Barz \cite{paesani_scheme_2021, chin_heralded_2024, bhatti_heralding_2024}. 
The scheme laid out by Chin \textit{et al.}~uses a method based on single-boson subtraction operators \cite{chin_boson_2024}, which they implement in their circuits through `subtractors'. 
Their optimised subtractor consists of a half-wave plate, followed by a polarising beam splitter. 
One output path of the polarising beam splitter is then routed to another half-wave plate and polarising beam splitter, with detectors on each of the outputs. For an $d = 2$, $N$-partite GHZ state, $2N$ input photons are needed. 
The photons are in $H,V$ polarisation states, and are incident on half-wave plates and polarising beam splitters before routing to $N$ copies of the optimised subtractors. 
If we herald one photon on each of the optimised subtractors, the output state is a  $d = 2$, $N$-partite GHZ state. 
The success probability (without feed-forward) is $p_{\text{success}} = \frac{1}{2^{2N}}$. 
Including feed-forward, this can be increased to $p_{\text{success}} = \frac{1}{2^{2N-1}}$.

The scheme laid out by Paesani \textit{et al.}~\cite{paesani_scheme_2021} requires $m$ deterministic input photons, with each photon encoding a qudit of dimension $d$. 
The photons are incident on an $m \times m$ DFT linear optical circuit, and the desired outcome is heralded by $m - N$ photons being detected on the top mode, with vacuum in the other heralding modes. 
The scheme finds that any $N$-photon, $d$-dimensional GHZ state can be generated solely with linear optics.
However, the heralding requirement on the top mode requires PNR detection, and the scheme requires large numbers of input photons, along with large interferometric schemes. 
This means an experimental realisation may be difficult at present. 
For example, using this method, a 3-photon 2-dimensional GHZ state generator requires 25 input photons and PNR detection of 22 photons at one output. Moreover, the associated success probability is approximately $10^{-10}$. 

A further scheme for generating any $N$-photon GHZ state is provided by Bhatti and Barz in \cite{bhatti_heralding_2024}. 
This uses the same SMS structures as in their Bell state scheme, as outlined in Sec.~\ref{sect:BSG}. 
For a 2-dimensional GHZ state with an even number of photons, $N_e$, $2N_e$ input photons are required. 
On the odd input modes of each 4SMS, identical input photons are in some state $\ket{\mu}$. 
On the even input modes of each 4SMS, identical photons, in some state $\ket{\eta}$ are input. 
Threshold detectors can be used to detect a single photon in the third output mode of the first 4SMS and in output mode 2 of the last 4SMS, both in the $\pm$ basis. 
We also detect one photon in each of the input states from each 2SMS output. 
This heralds a $N_e$-photon GHZ state with a success probability $p_{\text{success}} = \left( \frac{1}{2} \right)^{2 N_e - 1} $. 
This scheme can also be adapted for a GHZ state with an odd number of photons by including an additional detector. 
This gives a success probability of $p_{\text{success}} = \left( \frac{1}{2} \right)^{2 N_o} $. 

Bhatti and Barz's scheme for creating high-dimensional GHZ states \cite{bhatti_heralding_2024} is similar to their scheme for high-dimensional Bell states as laid out in Sec.~\ref{sect:BSG}. 
For $d$ dimensions, $4d$ input photons are required to create an $N = 3$ GHZ state, with one photon on each mode. 
Here, the optical circuit again comprises of 4SMSs, however PNR detectors at the final SMS structures herald the creation of a $d$-dimensional GHZ state when 4$d$ - 3 photons are detected. 
This occurs with a success probability of $p_{\text{success}} =  \frac{d \times 3^{d-1}}{2^{5d - 3}}$. 
This can be increased by using the bleeding method laid out in Sec.~\ref{sub_sec:multiplexing}. 
With bleeding, the success probability becomes  $p_{\text{success}} =  \frac{d \times 3^{d-1}}{2^{3d - 1}}$.

Another scheme to generate $N$-photon, $d = 2$ GHZ states that can also benefit from bleeding is found in  \cite{bartolucci_creation_2021}. 
The GHZ state generation circuit presented here is based on the circuit from \cite{Varnava2008}. 
Here, for an $N$-photon GHZ state, we require $2N$ input photons. 
A series of unit cells of interferometry are then carried out, which are referred to as `primates'. 
For each primate, a pair of modes in the interferometer and two additional modes with no input photon are used, and two of the modes are sent to a pair of detectors. 
This primate is repeated for each pair of modes in the interferometer. 
If there is one click on each pair of detectors at the output of each primate, this heralds the creation of a $d = 2$, $N$-photon GHZ state with success probability $p_{\text{success}} =  \frac{1}{2^{2n - 1}}$. 
Using bleeding here increases the success probability to $p_{\text{success}} =  \frac{1}{2^{n - 1}}$. 
An overview of these GHZ schemes is provided in Table \ref{table:ghz_schemes}.

\begin{table*}[htb]
\begin{tabularx}{\textwidth}{*{6}{| >{ \raggedright\arraybackslash}X}{|>{\raggedright\arraybackslash}X |}}
   \toprule
Scheme reference & Number of input photons $N$ & Number of modes & Success probability & Detector type\\ 
   \midrule
\cite{Varnava2008} & 6 & 12 & $1/64 \approx 1.562\% $  & Threshold\\
\hline
\cite{gubarev_improved_2020} & 6 & 10 & $1/54 \approx  1.852\% $  & Threshold\\
\hline
\cite{Walther2007} & 12 & 24 & N/A  & Threshold\\
\hline
\cite{Niu2009} & 10 &  16  & $1/16 \approx 6.25\% $\footnote{This number assumes that the input state is already available with certainty. In practice, the generation of the input state is also probabilistic and affects the overall generation rate. }  & PNR\\
\hline
\cite{Krenn2021} & 10 & 13 &  N/A  & Threshold \\
 \hline
\cite{chin_heralded_2024} & 6 & 12 & 1/64 without feed-forward, 1/32 with feed-forward & Threshold \\
\hline
\cite{paesani_scheme_2021} & 25 & 25 & $\sim10{^{-10}} \approx 10^{-8}\%$ & PNR\\
\hline
\cite{bhatti_heralding_2024} & 8 & 8 & $1/64 \approx 1.562\%$& Threshold\\
\hline
\cite{bartolucci_creation_2021} & 6 & 12 & $1/32 \approx 3.125\%$ & Threshold \\
   \bottomrule
\end{tabularx}
\caption{Comparison of existing schemes for heralded GHZ state generation, alongside their success probabilities and the type of detection required. For schemes that can generate $N$-partite GHZ states, the $N$=3 case is considered for comparison.}
\label{table:ghz_schemes}
\end{table*}

\subsubsection{GHZ state generation experiments}

Because of the serious step-up in experimental requirements compared to heralded Bell state generation, the first experimental demonstrations of heralded GHZ states have only taken place in recent times.
Three teams---a team associated with Quandela \cite{maring2024Pub}, Jian-Wei Pan's group in China \cite{chen2024pub}, and Philip Walther's group in Austria \cite{cao2024Pub}---all published experimental implementations of heralded GHZ generation in 2024.
All three experiments realised schemes that require six single photons as inputs.
Furthermore, the experiments all obtained photons from a quantum dot coupled to a microcavity.
The quantum dot emitted a train of single photons that were demultiplexed from time to path.
In terms of photon-source metrics, Maring \textit{et al.}~\cite{maring2024Pub} report a single-photon number purity of $>0.99$ and indistinguishability of $0.928\pm 0.011$, Chen \textit{et al.}~\cite{chen2024pub} report a single-photon purity of $0.974(6)$ and indistinguishability of $~0.87$, while Cao \textit{et al.}~\cite{cao2024Pub} report a single-photon purity of $0.981\pm 0.003$ and indistinguishability of $0.923\pm 0.009$.

Key differences between the approaches of the experiments are that Cao \textit{et al.}~\cite{cao2024Pub} implemented the 12-mode scheme of \cite{Varnava2008} with polarisation- and path-encoded photons in bulk-optics, Maring \textit{et al.}~\cite{maring2024Pub} implemented the same scheme but with path-encoded photons \cite{Li2015} in an integrated circuit at 928 nm, whereas Chen \textit{et al.}~\cite{chen2024pub} realised the 10-mode circuit \cite{gubarev_improved_2020} with path-encoded photons that had been frequency-converted to 1550 nm.
Both of the chip-based experiments \cite{maring2024Pub,chen2024pub} used fully programmable silicon nitride integrated circuits.
None of the experiments used PNR detectors, but the experiments of \cite{chen2024pub} and \cite{cao2024Pub} gained pseudo-photon-number-resolution by using beam splitters to divide detection modes into two modes incident on two detectors.

The reported fidelities between an ideal GHZ state and the three-photon states obtained when conditioning on the heralding signal are $0.82\pm 0.04$ in Maring \textit{et al.}'s experiment \cite{maring2024Pub}, $0.573\pm 0.024$ in Chen \textit{et al.}'s experiment \cite{chen2024pub} and $0.7278 \pm 0.0106$ in Cao \textit{et al.}'s experiment \cite{cao2024Pub}.
These fidelities are after postselecting on having three photons in the target modes, with the values not accounting for heralded events with zero to two photons in the target modes.
These events stem from runs when (a) fewer than six photons were injected into the circuit; when (b) photons that should have exited in the target modes are lost; or when (c) more than three photons exited in the heralding modes with only three detected, such that non-heralding events were misread as heralding events. 

\subsection{Other states} 
\label{subsect:other_states}

\subsubsection{Non-maximally entangled pure states} 
Whilst the aforementioned NOON, GHZ and Bell states are all maximally entangled states, heralded generation schemes can also be used for other, non-maximally entangled states, such as in the aforementioned schemes \cite{Ra2015,Vergyris2016}. Consider the dual-rail encoded two-qubit state, expressed in the Fock basis:
\begin{equation}
    \ket{\Phi(\alpha)} = \text{cos}(\alpha)\ket{0101}_{F} + \text{sin}(\alpha)\ket{1010}_{F},
\end{equation}
where $\alpha$ determines the degree of entanglement and varies between $0$ and $\pi / 4$ (by setting $\alpha = \pi / 4$, we recover a Bell state).

Fldzyhan \textit{et al.}~\cite{fldzhyan_compact_2021} consider both a 6-mode and 5-mode scheme to generate this state, with different heralding strategies. 
The six-mode linear optical circuit uses four photons input into the top four modes, and the bottom two modes support heralding. 
The successful creation of $\ket{\Phi(\alpha)}$ in the top four output modes is heralded by a single photon being incident on each of the detectors in the bottom two output modes. 
The five-mode linear optical circuit also requires four photons in the top four input modes, however creation of the state $\ket{\Phi(\alpha)}$ in the four top modes is heralded by PNR detection of two photons in the bottom heralding mode. 

Fldzyhan \textit{et al.}~\cite{fldzhyan_compact_2021} simulate these circuits for varying values of $\alpha$ and find the success probability of generating the output state.
The circuits realising the arbitrary unitary matrices with two heralding modes are more efficient than those with a single heralding mode at lower values of $\alpha$. 
As a result they are more efficient for weakly entangled state generation.
However, at higher values of  $\alpha$, the scheme with a single heralding mode is more efficient. 
For example, for the maximally entangled Bell state, the scheme with one heralding mode has a success probability of $p_{\text{success}} = 1/9$, whereas the scheme with two heralding modes has a success probability of $p_{\text{success}} = 2/27$.
Whilst a physical explanation is yet to be proposed, it highlights how the heralding scheme needs to be tailored to the desired state to achieve optimal success probability.

\subsubsection{Werner states}

Werner states are bipartite quantum states that are invariant under any unitary of the form $U \otimes U$ \cite{werner_quantum_1989}. For a two-qubit Werner state, this can be written as a convex combination of the maximally mixed state and a Bell state, 
\begin{equation}
    W_{AB}^{(\lambda, 2)} = \lambda \ket{\Psi^-} \bra{\Psi^-} + \frac{1-\lambda}{4} \mathbb{I},
\end{equation}
where $\lambda \in \left[0, 1 \right]$ is
the relative weight of the Bell state, and $\mathbb{I}$ is the $4\times4$ identity matrix. 

Methods for the heralded generation of Werner states can take inspiration from the schemes laid out in \cite{liu_experimental_2017, villegas-aguilar_nonlocality_2023}. Here, mixed states are demonstrated by using SPDC to create a Bell state, and then one photon of the pair is sent via a controllable probabilistic depolarising channel,  creating a temporal delay that couples polarisation and time degrees of freedom.
When tracing over the time degree of freedom, this creates a state locally equivalent to the Werner state, but the method relies on postselecting the SPDC source and the probabilistic recombination of the two photon paths. Adaptation of this method, however, could allow for the generation of heralded Werner states.
Firstly, the initial Bell state can be generated in a heralded manner, as described in Sec.~\ref{sect:BSG}. 
Fast switching can then be employed to turn the postselected recombination of photon paths into deterministic recombination. 
Alternatively, probabilistic recombination could be heralded by detecting vacuum on one of the output modes of the depolarising channel. 

\subsubsection{W states} 
For three qubits, the W state is the complementary state to the GHZ state where the measurement of one qubit leaves the remaining pair entangled \cite{dur_three_2000}. A W state for $N$ qubits can be written as
\begin{equation}
    \ket{W_N} = \frac{1}{\sqrt{N}} \left( \ket{100...0} + \ket{010...0} + ... + \ket{00...01} \right).
\end{equation}

A theoretical scheme for N-photon W states is suggested by Chin \textit{et al.}~in \cite{chin_heralded_2024}, which is devised using a linear quantum graph and bosonic subtraction, similar to that used in their scheme for N-partite GHZ state described in Sec.~\ref{subsec:GHZstates}. 
To generate the W state, their circuit uses $2N$ photons of $H,V$ polarisation incident on a half-wave plate and polarising beam splitter, and one ancillary $H$ photon incident on a half-wave plate, along with an $N$-partite port. 
These photons are then routed to $N + 1$ subtractors. $N$ of these are combinations of half-wave plates and polarising beam splitters, where the bottom mode of the polarising beam splitter is incident on a beam splitter, with both outputs of the beam splitter leading to detection. 
The remaining subtractor is a set of $N$ polarising beam splitters where the bottom mode of each polarising beam splitter is routed to an $N$-partite port, with detectors on each output of the port. 
We then herald on one detector in each of the $N+1$ subtractors detecting one photon, routing the remaining photons to a final set of polarising beam splitters to generate an N-photon W state.
This results in a success probability of $p_{\text{success}} = \frac{1}{N 2^{2N+1}}$ without feed-forward. With feed-forward, this is increased to $p_{\text{success}} = \frac{1}{2^{2N}}$.

\subsubsection{Larger states via fusion gates}
\label{subsubsect:fusion_gates}

\begin{figure}
    \centering
    \includegraphics[width=0.75\columnwidth]{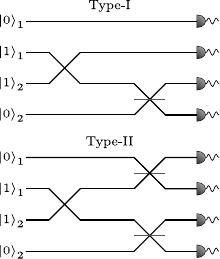}
    \caption{Type-I and type-II fusion gates. Both gates act on a pair of photonic qubits where one and both qubits are measured in type-I and type-II, respectively. The Type-II fusion gate corresponds to a linear-optical Bell state analyser.}
    \label{fig:fusion_gates}
\end{figure}

For some applications, there is a need for large entangled states.
Fusion gates form a group of important linear optical gates, which can take smaller entangled states and \textit{fuse} them together in a heralded manner. 
When first introduced in \cite{browne2005resource}, these were grouped into two types, distinguished by the number of qubits measured. 
We depict each in Fig.~\ref{fig:fusion_gates}.
Both fusion gates are probabilistic, and it has been shown (see e.g. Refs.~\cite{calsamiglia_maximum_2001, PhysRevA.84.042331, PhysRevLett.113.140403}) that, without additional ancilla photons, fusion gates are limited to a success probability of $1/2$.
Type-I fusion has the benefit of reduced resources, as it only measures one of the two qubits, leaving the remaining qubit entangled with all other unmeasured qubits.
However, if this qubit is lost, there is no way to detect this error.
In contrast, both qubits are measured in a type-II fusion, meaning if one is lost, the error can be heralded. 
Hence, loss detection is built into the type-II fusion gate, unlike in type-I fusion.
To grow larger entangled states,  Bell states can be used with type-I fusion gates. 
When using type-II gates, net growth can only be achieved with three-photon GHZ or larger entangled states as the initial resource. 
In applications requiring large entangled states, both gates can be combined to leverage the smaller resource requirements of type-I fusion, along with the loss tolerance of type-II.

\subsection{Increasing success probability}
\label{sub_sec:multiplexing}

Whilst using ancillas to herald a target state of interest allows the state to be generated in an event-ready manner, as more photons and larger linear optical circuits are required, the overall success probability tends to decrease. 
There are several approaches, including using further ancilla photons, or measurement results along with feed-forward processing, to improve the success probability of state generation circuits.

\subsubsection{Boosting}

Boosting include ancilla photons to improve the success probability of schemes. 
For example, a standard Bell state measurement (BSM), corresponding to a type-II fusion gate, has a success probability of 50\%, as it can differentiate two out of the four possible Bell states. 
Different boosted BSM schemes have been devised, for example, a boosted BSM which includes an additional entangled ancilla pair that increases the success probability of the BSM to 75\% is shown in \cite{Ewert20173/4}. 
Whilst this can be beneficial in increasing the success probability, it may require additional experimental overheads, such as PNR detection and a higher number of input photons. 

This can be used in some of the aforementioned heralded entanglement schemes, for example, the 4P8M Bell state generator success probability can be increased to  $p_{\text{success}} = 13/64$ with two additional ancillary photons \cite{bartolucci_creation_2021}. 
Boosted BSM can also be beneficial to fusion gates, increasing both the success probabilities of the type-I and type-II fusion gates discussed in the previous section. 
By using a Bell state as ancillary resources here, the success probability can be increased to  $p_{\text{success}} = 3/4$. 

An experimental demonstration of boosted BSM has recently been shown in \cite{hauser2024boostedbellstatemeasurementsphotonic}. 
Here, two input photons and an entangled ancilla pair are input on a $4\times4$ fibre multiport splitter, implementing a $4\times4$ DFT. 
The experimental BSM is found to have a success probability of $0.693 \pm 0.003$.
This can have an impact on \textit{fusion-based quantum computing}, as described in Sec.~\ref{QC_subsect}. 
The photon-loss threshold --- or the probability above which the logical error rate begins to decrease as the size of the system increases --- can be affected by using a boosted BSM.
In \cite{hauser2024boostedbellstatemeasurementsphotonic}, the authors simulate how their BSM would improve the photon-loss threshold for the six-ring resource state example, finding that the photon-loss threshold here is almost three times higher than for the non-boosted case, making their proposal more robust to loss. 

\subsubsection{Locally equivalent output patterns}

A simple approach makes use of the fact that different heralding patterns can lead to states that are locally equivalent to the desired target state, that is, states that can be transformed into the desired target state by local transformations.
For example, in the case of the 4P8M Bell state generator, each heralding pattern with two photons in different heralding modes can be useful. 
Depending on which of the six possible heralding patterns is observed, one of the four Bell states is generated, each with a success probability of $1/32$. 
If we can feed-forward single-qubit rotations required to convert each of these into our desired state, we can increase the success probability to $3/16$. 
There are also output patterns with two photons in the same heralding mode that lead to non-maximally entangled, W-type states in the Fock basis. 
A linear optical network can be applied here, forming a Schmidt decomposition to the following state:
\begin{equation}
    \sqrt{\frac{3}{4}} \ket{2000}_{F} + \sqrt{\frac{1}{12}} \ket{0200}_{F} + \sqrt{\frac{1}{12}} \ket{0020}_{F} + \sqrt{\frac{1}{12}} \ket{0002}_{F}.
\end{equation}

To reduce the first term, an ancillary vacuum mode is coupled to the first mode, and heralded on detecting the vacuum. 
A linear optical network can then be applied to take the resultant state back to a dual-rail encoding, to form a maximally entangled Bell pair. 
Through these distillation methods, the success probability for the 4P8M case is improved to $p_{\text{success}} = 1/4$ \cite{bartolucci_creation_2021}. 

\subsubsection{Bleeding}\label{subsubsect:bleeding}
When a measurement is made on the heralding modes, this can be considered as applying a measurement operator, $\hat{M}$. 
These measurements gives the operator of the entire scheme on a single mode, namely:
\begin{align} 
\hat{M}_{(0)} = 2^{-\hat{n}/2}, 
\label{eqn:operators_bleeding}
\\ 
\hat{M}_{(1)} = 2 \hat{A} ^{-\hat{n}/2}.
\label{eqn:operators_bleeding2}
\end{align}
where $\hat{A}$ is the annihilation operator on one of the target modes, and $\hat{n}$ is the photon number operator of the target mode \cite{bartolucci_creation_2021}. 

\begin{figure}
    \centering
    \includegraphics[width=\linewidth]{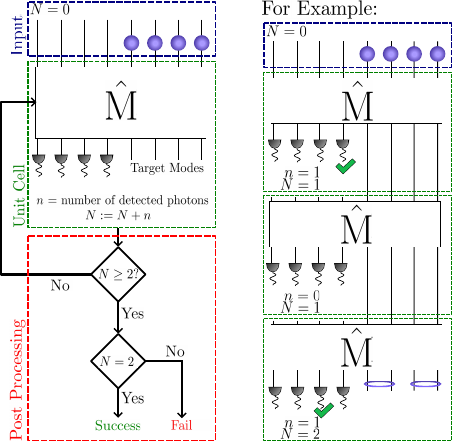}
    \caption{Figure showing the stages of the bleeding protocol for the example of the 4P8M Bell state generator. The outcomes from the detectors on the heralding modes are used to indicate if the scheme has been successful. An example is also included to show how the algorithm applies to different detector patterns.}
    \label{fig:bleeding_scheme}
\end{figure}

Considering for example the 4P8M Bell state generator circuit, if photons are measured on the first two of four heralding modes, this can be described using $\hat{M}_{(1100)}$, which can be constructed from  Eqs.~(\ref{eqn:operators_bleeding}) and (\ref{eqn:operators_bleeding2}).
The same operator can also be achieved by applying two single photon detection operators, applying $\hat{M}_{(1000)}$, followed by $\hat{M}_{(0100)}$, as shown in Fig.~\ref{fig:bleeding_scheme}.
Also, applying $\hat{M}_{(0000)}$ does not change the state produced on the target modes, and it commutes with the single photon detection operators. 

If either one or no photons are detected, the operators are repeatedly applied until two photons are detected, as shown in Fig.~\ref{fig:bleeding_scheme}. 
If more than two photons are detected on the output modes, the algorithm fails and no Bell state is produced. 
This increases the success probability for the 4P8M Bell state generator to $p_{\text{success}} = 1/2$. This can be further increased by using distillation, a process by which copies of the entangled state are transformed into a smaller number of maximally entangled states. The success probability is $p_{\text{success}} = 2/3$ when distillation is used, for two photons incident on the same detector \cite{bartolucci_creation_2021}. 
Bleeding relies on switching between subsequent stages of the linear optical circuit based on previous measurement outcomes. 
This feed-forward requires the repeated application of optical switches, which can also be lossy. 

\subsubsection{Multiplexing} \label{multiplex}

Once the circuit success probability has been optimised with approaches such as the ones above, we can use a technique known as multiplexing to increase the success probability further. 
Here we take many probabilistic events and attempt them in parallel. 
The processes can be duplicated in any photonic degree of freedom, for any probabilistic process. 
If the probability of a single process being successful is $p_\mathrm{success}$, then the probability that at least one of $N$ processes (or attempts) is successful is given by 

\begin{equation}
    p_\mathrm{tot} = \left(1- \left(1 - p\right)^N\right).
\end{equation}

By increasing the number of trials, we can drive this probability arbitrarily close to unity. 
Whilst the success probability for one event is now very high, the successful event will be in a random, but known, set of modes. 
Therefore, if we wish to use multiple devices together, we require a method to process the heralding events and then switch the successful event into a desired set of modes. 
For example, where multiple physical copies of the process are used, shallow-depth $m$-to-$n$-mode switching networks have been developed \cite{bartolucci2021switch}. 
If a single copy of a process is repeated many times, then quantum memories, either solid state \cite{nunn2013enhancing} or optical fibre \cite{mendoza2016active}, can be used to delay the successful event. 

To date, multiplexed single photons have been experimentally demonstrated using time \cite{adcock2022enhancement,kaneda2015time}, space \cite{francis2016all,collins2013integrated}, frequency \cite{joshi2018frequency} and orbital angular momentum \cite{liu2019multiplexing}.
While only single photon multiplexers have been demonstrated, the procedure can be adapted for any heralded probabilistic process \cite{bartolucci2021switch}.

\subsection{Methods for finding circuits}

\subsubsection{Motivation}
In the previous sections, we described known optical circuits to generate certain entangled states. 
Here, we consider more general techniques for finding circuits to create a given target state. 
In general, it is sufficient to identify a unitary matrix representing a linear optical transformation, which takes an input state that is easier to prepare experimentally into an output state that corresponds to a desired entangled target state conditioned on a certain heralding pattern (cf.~Fig.~\ref{fig:postAndHerEntanglement}b). 
Then it is well known how any unitary matrix can be converted into an optical circuit using sequences of beam splitters and phase shifters \cite{reck_experimental_1994, clements_optimal_2016}. 
There are two main approaches to finding unitary matrices to achieve a desired transformation. 
The analytical approach provides a rigorous method to prove the existence of and find a suitable unitary. 
Additionally, when the unitary is non-unique, the success probability can be optimised. 
However, this approach is usually inefficient and, thus, computationally impossible to perform above a certain system size. 
The other approach is based on numerical optimisation. 
Since the optimisation task is usually nonlinear and nonconvex, it is challenging to prove the optimality of a given solution, with heuristics used to explore the solution space and propose good candidate circuits. 

\subsubsection{Analytical approach}
VanMeter \textit{et al.}~proposed a method that relies on the use of Gr\"{o}bner bases to find suitable candidate unitary matrices, and on convex optimisation to identify the candidate matrix with the maximum success probability \cite{vanmeter_general_2007}. 
The Gr\"{o}bner basis method is a well-established technique in algebraic geometry \cite{cox_grobner_2015}. 
Typically, it can be used to solve a system of multivariate polynomial equations by rewriting it in a triangular shape, similar to Gaussian elimination. 
The problem is then greatly simplified since the system of equations is cascaded, such that each equation can be solved by finding roots in only one variable. 

VanMeter \textit{et al.}~describe how the problem of finding a suitable unitary to transform a given input state into a heralded target state can be cast as a system of multivariate polynomial equations, thereby opening a solution pathway by the Gr\"{o}bner basis method. 
To this end, they highlight that any state in linear optics can be represented as a polynomial in the creation operators of the optical modes. 
Therefore, any input state to the circuit, output state after the circuit, and target state can be written in terms of multivariate polynomials. 
The constraint that the transformation of the input state, followed by conditioning on the heralding pattern, must match the target state, yields a system of equations constraining a suitable transformation matrix. 
It is difficult to reconcile the unitarity condition of the transformation matrix with the algebraic geometry framework without foregoing valuable tools of the framework. 
Therefore, the authors removed the unitarity condition on the transformation matrix, and the unitary property is restored at the end by embedding a non-unitary solution as a block of a larger, unitary matrix. 
This is achieved by adding additional ancilla modes populated with vacuum at the input, and heralding on the vacuum in the associated output modes. 
When solutions to the state generation problem exist, they are non-unique in general. 
If the Gr\"{o}bner basis technique provides one or a set of solutions to the system of polynomial equations, there are a few degrees of freedom by which each solution can be rescaled to give other solutions, giving rise to equivalence classes. 
Then it is possible to perform optimisation within the equivalence class to find the representative associated with the highest success probability. 
VanMeter \textit{et al.} describe how this optimisation can be performed by using a hierarchy of semidefinite programmes that can be solved efficiently using standard convex optimisation techniques \cite{boyd_convex_2004}. 

\begin{figure}
    \centering
    \includegraphics[width=\columnwidth]{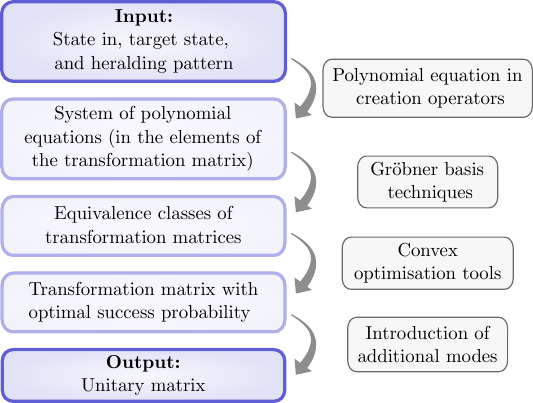}
    \caption{Overview of key steps in the analytical method introduced by VanMeter \textit{et al.}~\cite{vanmeter_general_2007}, starting from the method's inputs that define the state generation problem (top), and resulting in the unitary matrix that describes the linear optical transformation to be implemented (bottom).}
    \label{fig:grobner}
\end{figure}

Overall, VanMeter \textit{et al.}~propose a systematic approach to answer the question of the existence of a quantum state generator and to find the circuit with optimal success probability. 
The main steps of this procedure are presented in Fig.~\ref{fig:grobner}. 
An important caveat is that the worst-case complexity of finding Gr\"{o}bner bases is doubly exponential in the number of variables \cite{bardet_complexity_2015}. 
Even though in many cases the algorithm performs better than this, the method can still become impractical beyond small system sizes. 

\subsubsection{Numerical methods}
The use of numerical methods to aid the design of optical circuits has increased significantly in recent times. 
Typically, the overall solution space---the ensemble of valid unitaries in a fixed dimension---is parameterised. 
A cost function is then formulated in this space and optimised, to take into account the fidelity with the target state, the success probability, and the complexity of the circuit in terms of the number of optical elements \cite{fldzhyan_compact_2021, gubarev_improved_2020}.  
Alternatively, in \cite{gubarev_fock_2021} Gubarev proposed another formulation to directly enforce a unit fidelity, instead of including the fidelity in the cost function. 
Gubarev relies on the analytical criteria of \cite{garcia-escartin_method_2019} to enforce that the transformations are realisable using linear optics. 
The formulation results in a modified line search method that maintains the unit fidelity and unitary constraints, hence only the success probability needs to be optimised. 

These numerical methods typically rely on gradient descent techniques. 
For convex problems, these are known to converge to the optimal solution. 
However, here the problem is, in general, nonconvex, so gradient descent techniques are not guaranteed to find the optimal solution and may get stuck in local optima. 
To manage this issue, optimisation is usually repeated many times with a random initial point to better explore the parameter space. 
Numerical searches were successfully used to identify circuits to create Bell state generators with 4P6M, 4P5M, 4P5M with feed-forward, as shown in Table \ref{table:bsg_schemes}, and GHZ generators with 6P10M, as reported in Table \ref{table:ghz_schemes}. 
Additionally, the approach of \cite{gubarev_fock_2021} revealed that the 4P6M scheme from \cite{carolan_universal_2015} is suboptimal, and the success probability can be slightly improved. 

\section{Applications of heralded entangled states} \label{applications}

In the previous sections we have introduced heralded state generation and various schemes for its realisation. 
Here, we consider how heralded states can be used in different quantum technologies, focusing on applications in computing, communication and metrology.

\subsection{Quantum computing} \label{QC_subsect}

Small entangled states form the backbone of most modern envisagings of a photonic quantum computer. 
In this section, we review modern approaches to LOQC and show how the circuits and ideas discussed so far play a vital role in their construction. 

\subsubsection{Measurement-based quantum computing}

\begin{figure}[ht]
     \centering
     \includegraphics[width=0.5\textwidth]{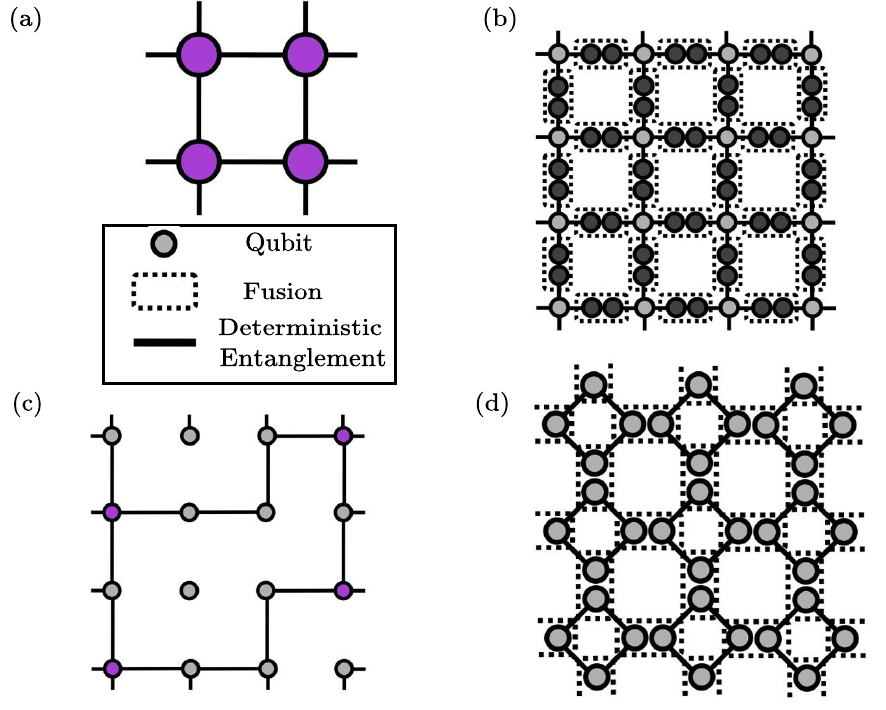}
    \caption{Percolation-based linear optical quantum computing (LOQC) and fusion-based quantum computing (FBQC) examples. (a)--(c) illustrate the use of renormalisation to overcome probabilistic entanglement links. An example desired entanglement structure is shown in (a). (b) shows a layout of three-qubit GHZ states and fusion gates, which can be used to generate the same structure. Light grey indicates qubits that will be part of the final entangled state and dark grey indicates qubits that are measured in fusions to generate the entanglement. (c) shows certain fusion gates having failed, indicated by a lack of an edge between nodes. The purple nodes represent the original nodes in (a) with links between them going through extra qubits. (d) shows an example FBQC network. Again, we start with small deterministically entangled states, but now all qubits are measured in fusion gates. (d) is reconstructed from  \cite{bartolucci2023fusion}.}
    \label{fig:loqc_fig}
\end{figure}

As photonic qubits travel and have probabilistic two-qubit gates, gate-based quantum computing, where a single qubit must undergo many gates, is impractical.
In fact, initial LOQC architectures adopting this model resulted in very large overheads to overcome these issues \cite{knill_scheme_2001}.  
Modern architectures tend to favour an alternative but equivalent model of computation, often called \textit{measurement-based quantum computing} (MBQC). 
In this picture, an initial large entangled state known as a cluster state, is used, and the computation proceeds by single-qubit measurements.
A quantum algorithm is implemented by a sequence of single-qubit measurements, with feed-forward of measurement outcomes conditioning subsequent measurements. 
Many of the initial approaches to photonic MBQC still attempted to make the entangling links quasi-deterministic using repeat-until-success schemes, thus incurring large overheads \cite{nielsen2006cluster,browne2005resource}.
Then it was shown that ideas from percolation theory could be used to implement MBQC with probabilistic gates \cite{kieling2007percolation,gimeno2015three,pant2019percolation}.
The main idea behind these percolated architectures was that so long as the success probability of the entangling gates was high enough, there would be a path for quantum information to propagate from one end of the cluster to another. 
This behaviour exhibits a threshold where, above a certain edge probability, increasing the size of the cluster state increases the probability that there exists a set of edges spanning the cluster.
This can then be combined with a technique known as renormalisation whereby, so long as we know which edges were successfully created, we can produce the same entanglement structure as the desired lattice, as shown in Fig.~\ref{fig:loqc_fig}.
Initially, these architectures required large, deterministically generated entangled resource states, with only a few probabilistic links joining resource states \cite{kieling2007percolation}. 
However, in \cite{gimeno2015three}, it was shown that only GHZ states and heralded fusion gates are sufficient for universal quantum computing.

\subsubsection{Fusion-based quantum computing}\label{subsubsec:fbqc}

The most modern LOQC architectures take an approach similar to MBQC. 
However, instead of generating a large entangled state and then using single-qubit measurements, fusion-based quantum computing (FBQC) uses multi-qubit measurements acting on small entangled states generated beforehand. 
The FBQC approach allows for the generation of large-scale entanglement in \textit{fusion networks}, which can then be used to implement algorithms. 

Two-qubit fusion operations can be achieved using BSMs, which, as we have seen, can be implemented with only linear optics, along with the capability of boosted success probabilities through the consumption of extra ancilla photons. 
Larger n-qubit joint measurements can also be constructed from these BSMs \cite{pankovich2024high}. 
A key caveat here is that as percolation is not used here, error-correcting codes must be used to account for erasure errors stemming from the finite failure probability of photonic fusion gates \cite{bartolucci2023fusion}. 
The number of components each photon passes through, known as the optical depth, is independent of the computation size. 
Only the number of resource state generators required increases. 
Beyond this, the optical depth can be kept relatively short, as in principle once created by the resource state generator, the photons pass through a few beam splitters before being measured. 
Optical delays can be used to decrease the spatial extent of the network, but this is fixed throughout the computation \cite{bombin2021interleaving}. 
The current resource states proposed in the literature are the 4-photon GHZ states and 6-qubit ring states \cite{bartolucci2023fusion}, Bell states \cite{pankovich2024high}, and linear cluster states \cite{paesani2023high}. 
Repetition encoding has also been proposed. 
Here each photon is represented as an entangled state ($\ket{0} \rightarrow \ket{0}^{\otimes N}, \ket{1} \rightarrow \ket{1}^{\otimes N}$) \cite{bartolucci2023fusion,pankovich2024high}. 
However, approaches based purely on linear optics aim to build the desired resource states from smaller \textit{seed states} \cite{bartolucci_creation_2021,bartolucci2021switch,pankovich2024flexible}. 
These approaches combine such as Bell and 3-photon GHZ states, constructing larger entangled states through probabilistic fusion gates and multiplexing.
The requirement for multiplexing in these approaches demonstrates the importance of heralded generation of the seed states.

\subsection{Quantum communication}{\label{QuantCom}}

As with any application of quantum information, loss poses a significant challenge within quantum communication. 
This issue becomes even more pronounced in long-distance quantum communication, where loss not only leads to the failure of quantum information protocols but also severely limits the transmission distance due to its exponential scaling with distance. 
For instance, when transmitting single photons over 100 km of standard telecommunication fibre, only one photon out of 100 typically survives. 
In this context, we highlight how heralding techniques can be employed to mitigate loss and enhance the performance of communication tasks.

\subsubsection{Entanglement swapping: entanglement sharing between two distant parties}

In quantum networks, the ability to share entanglement between distant parties holds immense significance for large-scale quantum information processing and serves as a crucial resource for ensuring secure quantum communication~\cite{kimble2008quantum,lu2021quantum}. 
Entanglement swapping is designed to generate entanglement between two parties via the intermediate use of a middle station. 
The basic protocol works as follows~\cite{Pan1998}. 
Each party generates locally an entangled Bell state between a photon and another system kept in a quantum memory. 
Each photon is transmitted to a middle station which is then performing a BSM on the two photons. 
The outcome announced by the middle station determines if the operation has succeeded, in which case the remaining two systems in their memories are now entangled, even though they have never interacted directly. 

\begin{figure}[ht]
     \centering
     \includegraphics[width=0.6\columnwidth]{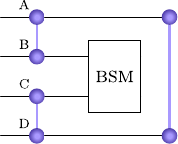}
    \caption{Entanglement swapping from the A-B and C-D pairs to the A-D pair. Particle A is entangled with particle B, and particle C is entangled with particle D. Half of each entangled pair is sent to the middle station, where they are projected onto a Bell state by a Bell state measurement (BSM). After this measurement, particles A and D become entangled, even though they have never directly interacted.}
    \label{fig:ent_swap_fig}
\end{figure}

This procedure is illustrated in Figure~\ref{fig:ent_swap_fig}. 
In this scenario, where the success is heralded by the central station, losses in modes B and C, including detection inefficiencies and the probabilistic nature of the Bell state measurement, can be disregarded. These losses lower the generation rate but do not influence the quality of the heralded generated state.
In photonic systems, where deterministic generation of entangled pairs is unfeasible, heralded schemes can be employed to obtain entangled states between parties A-B and C-D. 
These heralded schemes offer a practical approach to mitigating the challenges posed by probabilistic pair generation, such as double pair generation of A-B with nothing generated in C-D and vice versa, ensuring that only the intended number of photon pairs is retained.

Entanglement swapping can be employed in different communication protocols such as quantum repeaters~\cite{Azuma2024}, entanglement-based key distribution~\cite{Scherer2011long}, quantum secure direct communication~\cite{gao2004quantum} and quantum bidirectional secure direct communication~\cite{Chen_Yan_2007}.  
From an experimental point of view, entanglement swapping between photons~\cite{Pan1998}, atoms~\cite{Hofmann2012}, and ion traps~\cite{riebe2008deterministic} has already been realised. 

\subsubsection{Heralding to remove the detection loophole}
Heralded entanglement generation is a useful technique in tests of nonlocality between matter qubits to mitigate a critical issue called the \textit{detection loophole} \cite{brunner_bell_2014}. 
This issue arises from the fact that measurement devices do not always give a result, for instance in photonic systems when the photon has been lost during the transmission. 
An intuitive way of dealing with this problem is to ignore all the instances when the detectors do not click. 
However, this approach based on postselection of the conclusive outcomes relies on the assumption of fair sampling \cite{brunner_bell_2014}, which assumes that the observed statistics still faithfully represent the quantum system once restricted to the conclusive outcomes. 
However, in some adversarial scenarios, these statistics could be distorted due to an adversary blocking certain events not in their favour. 
This is always a possibility when the overall system efficiency is lower than a certain threshold, in which case this issue is commonly referred to as the detection loophole. 
The existence of this loophole is an argument to challenge the validity of certain security statements or fundamental results \cite{brunner_bell_2014}. 
Nevertheless, in photonics, it is possible to use more advanced postselection techniques to process the inconclusive outcomes \cite{liu_towards_2023}. 
In matter-based qubits, it is possible to use a heralded entanglement distribution scheme to remove the generation inefficiency and the losses resulting from detection inefficiency and the probabilistic nature of the Bell state measurement from the analysis. 
Upon announcement of a successful entanglement distribution, the measurement devices have to give a conclusive outcome. 
This heralding mechanism has been at the basis of some loophole-free Bell test experiments \cite{hensen_loophole-free_2015, rosenfeld_event-ready_2017}, and more recently of two of the three reported proof-of-concept experiments of device-independent quantum key distribution \cite{nadlinger_experimental_2022, zhang_device-independent_2022}. 
In all of these experiments, entanglement is generated locally between a matter qubit and a photonic qubit, and the photonic qubits are transferred to a middle station to perform entanglement swapping as described in the previous section. 
Successful entanglement swapping leaves the two matter qubits entangled, and ready for further processing. 
The inefficiency of the entanglement swapping due to the probabilistic nature of BSM, transmission losses, and detection inefficiency in the middle usually results in low generation rates, but all the unsuccessful events can be ignored without opening a detection loophole. 
Then the measurement of the local matter qubits typically benefits from the near-unit detection efficiency for this platform. 
The quality of the distributed entanglement can then be tested, and even certified using a Bell test \cite{brunner_bell_2014}. 
Different applications have different requirements on the quality of the entanglement. 
While a fundamental Bell test only requires a violation of a Bell inequality by several standard deviations, an application to device-independent quantum random number generation requires a high violation value, and applications to device-independent quantum key distribution additionally require a low error rate \cite{primaatmaja_security_2023}.

\subsection{Quantum metrology}{\label{QuantMet}}

 One of the applications where quantum resources can provide an advantage is metrology~\cite{giovannetti2004quantum, giovannetti2006quantum,giovannetti2011advances}, that is, the measurement of physical parameters. 
 Metrology often uses light as the probe, with the archetypal measurement task being phase estimation. 
 Since there are a plethora of cases where a physical effect of interest causes an effective path length difference, phase estimation has widespread applications, from gravitational wave detection \cite{LIGOGravW} to measurements of biological systems \cite{Taylor2014}. 
 A basic resource in quantum metrology is the number of probe systems used, in our case, the number of photons \cite{Polino_review2020}. 
 A quantum metrological advantage means achieving a precision that is impossible to obtain with classical resources and strategies, for a given number of photons. 
 Thus, in a fully rigorous analysis of a quantum metrology experiment, all the photons passing through the sample must be counted as resources.

A useful tool for designing metrological strategies is the Fisher information, which quantifies how much information about the parameter of interest is contained in the probability distributions underlying the measurement outcomes \cite{Polino_review2020}. 
Such strategy design can identify what quantum states to use as the probe, along with the choice of interaction and read-out. 

A few quantum states of interest for metrology exist that can be prepared deterministically if deterministic single-photon sources are available. 
Such deterministic preparation is possible for two-photon NOON states and for a general family of states called Holland-Burnett states \cite{Holland1993}, which are produced by a balanced beam splitter with equal Fock states incident on the two input ports. 
In the absence of deterministic single-photon sources, two-photon NOON states can be created probabilistically from SPDC in such a way that one either has the NOON state or vacuum (up to higher-order SPDC terms). 
This approach has been used to experimentally demonstrate a rigorous quantum metrological advantage \cite{slussarenko2017unconditional}. 
The above state preparation schemes have in common that no photons other than states of interest pass through the sample. 

\begin{figure}[ht]
     \centering
     \includegraphics[width=\columnwidth]{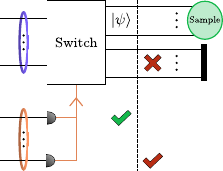}
     \caption{Selective use of successful state generation events for quantum metrology applications. A state generation circuit (not shown) outputs target modes (circled in purple) and heralding modes (circled in orange). A switch guides the target modes either to the sample or to a beam block (black bar), depending on a classical feed-forward signal (orange lines) based on the detection pattern on the heralding modes. In this way, the desired state $|\psi\rangle$ is guided to the sample when the correct heralding pattern is obtained (green tick); undesired states (red cross), which correspond to an incorrect detection pattern (red tick), are discarded.}
     \label{fig:metrology}
 \end{figure}

However, other states one might wish to use for metrology \cite{Polino_review2020} would generally require probabilistic preparation of the kind where unsuccessful generation attempts produce states other than vacuum. 
Postselection-based state preparation schemes leave no choice but to have the unsuccessful state generation attempts contribute to the resource count while providing sub-optimal information, thereby reducing the Fisher information and diluting or altogether negating the quantum advantage. 
Heralded state preparation, by contrast, unlocks the important ability to block these unwanted outputs: One can employ fast optical switching~\cite{bartolucci2021switch} conditioned on the heralding signal to only release the state of interest and block photons from reaching the sample whenever the state preparation is unsuccessful. 
This idea is illustrated in Fig.~\ref{fig:metrology}.

Fast switching has been realised in a metrology experiment that released single photons generated by SPDC after heralding their partner photons, thereby lowering the detector efficiency necessary for quantum-enhanced transmission measurements \cite{Sabines2017}. 
Although a number of experimental efforts used other heralded states for metrology \cite{ThomasPeter2011,thekkadath2020quantum}, there is an outstanding challenge to achieve a quantum advantage using entangled states.
Using optical switches to selectively release heralded entangled multi-photon states as a measurement probe and achieving a quantum advantage will be an exciting future prospect and a significant milestone in quantum metrology.
 
\section{Conclusion}

In contrast to other quantum information processing platforms that allow deterministic two-qubit gates, for discrete-variable photonic encoding, the preparation of general states with linear optics and photon detection necessarily takes on a probabilistic character.
There are two main approaches to identifying successful events in photonic state generation: postselection and heralding.

Postselected state generation schemes are typically easier to experimentally implement because they require fewer photons and modes.
Hence, postselected schemes have been the go-to approach for most proof-of-concept experiments, and have allowed the preparation of many different states such as high-dimensional Bell states \cite{Schaeff2015,Wang2017}, GHZ states with up to $12$ photons \cite{Bouwmeester1999,Pan2001,Zhao2004,Lu2007,Huang2011,Yao2012,Wang2016,Zhong2018} and up to three local dimensions \cite{Erhard2018}, up to $4$-photon W states, \cite{Eibl2004,Mikami2005,Tashima2010} and up to $6$-photon Dicke states \cite{Kiesel2007,Prevedel2009,Wieczorek2009}.
In fact, postselection is so widespread among experimentalists that the explicit mention of its use is often omitted in experimental literature because it is treated as the default.
However, the approach of postselection is fundamentally limited. Successful events are only identified after the state is destroyed through measurement, making it impossible to block unwanted states from the remainder of the optical circuit or to increase the success probability through multiplexing.
Moreover, the need to preserve the ability to postselect at the final photon detection stage imposes severe restrictions on the states that can be generated \cite{adcock_hard_2018} and the type of processing that can be carried out on the state.

Conversely, heralded state generation is scalable and does not suffer from these shortfalls. 
The measurement of a heralding pattern constitutes a signal of successful events, leaving the target state ready to be processed further or used without restrictions.
As heralded schemes are typically more demanding than postselected schemes in terms of resources, this approach is not yet as well developed.
However, the transition that is taking place from proof-of-concept to mature photonic technologies is now creating a strong impetus for heralded schemes.
This drive is exemplified by the recent experiments of three different teams demonstrating the heralded generation of three-photon two-dimensional GHZ states \cite{maring2024Pub,chen2024pub,cao2024Pub}.

Experimental imperfections can cause a deviation from the behaviour expected in theoretical state generation schemes. Photon loss is a major challenge in photonic quantum systems, impacting the fidelity and success of heralded state generation. Loss occurs through scattering, absorption, coupling inefficiencies, and detector limitations, leading to false positives (when heralding signals are observed without generating the desired state) and false negatives (discarding valid states due to missed heralding signals). Detector imperfections have a number of impacts: dead time reduces the generation rate; and high jitter and dark counts degrade the output state in addition to causing false-positive and false-negative heralding patterns, depending on the scheme. Additionally, photon distinguishability introduces decoherence, reducing interference quality and state fidelity. Multiphoton errors from probabilistic sources complicate heralding by introducing noise through extra photons in either target or heralding modes. Mitigating these issues requires optimised component design, gated detection windows synchronised with laser pulses, and photon-number-resolving detectors to enhance scalability and fidelity in quantum applications.

Heralded states provide a bedrock for quantum information technologies, from quantum computation to quantum communication and metrology. 
Perhaps their most pertinent role is as a fundamental resource in state-of-the-art discrete-variable quantum computing architectures such as FBQC \cite{bartolucci2023fusion}. 
Now that heralded Bell and GHZ state generation has been demonstrated, they need to meet the requirements for practical applications, with future focus on the creation of larger states. 
Transitioning to the generation of entangled states with more than three photons will allow for the construction of fusion networks comprising (logical) resource states such as (encoded) 4-GHZ and 6-ring states. 
Current photon loss threshold analysis for the FBQC platform suggests such encodings will be required for fault-tolerance in current experimentally achievable integrated photonic platforms~\cite{melkozerov2024analysis}. 
These considerations show us that there is still significant required progress in the context of heralded resource state generation for computational applications.
Quantum networking, the linking and communication of disparate quantum devices, can be enabled through heralded photonic schemes. 
One such example is the principle of photon-based entanglement swapping which forms the crux of proposals for quantum repeaters and quantum key-based communication. 
For quantum metrology schemes, heralding holds the important promise of switching mechanisms that can prevent undesired states, corresponding to failed generation attempts, from passing through a sample.

Heralded state generation stands to benefit from photonic technologies that are being developed for multiple purposes, including fast switches and feed-forward, quantum memories, quantum sources, and photon detectors.
To date, there are many open problems in the area of heralded state generation, including the generation of W states, high-dimensional Bell states, GHZ states with four or more photons, other computational resource states such as 6-ring states, multiplexing the generation of states comprising more than one photon, and the development of improved techniques for determining circuits.
The confluence of open problems, new photonic capabilities, and improved computing power, including artificial intelligence, is creating a fertile ground for upcoming advancements in heralded photonic state generation.

\section{Acknowledgments}
This work was supported by the Australian Research Council; N.T.~is a recipient of an Australian Research Council Discovery Early Career Researcher Award (DE220101082).  S.S.~acknowledges support from the Australian Government Research Training Program (RTP). F.G.~was supported partly by the Griffith University Postdoctoral Fellowship (GUPF\#58938).
This work was supported by the UK Research and Innovation Sciences and Technologies Facilities Council (UKRI STFC) and by the Engineering and Physical Sciences Research Council (EPSRC) Hub in Quantum Computing and Simulation (EP/T001062/1).
%
\bibliographystyle{apsrev4-1}

\end{document}